\documentclass[reprint, superscriptaddress, amsmath,amssymb, aps, floatfix]{revtex4-2}

\makeatletter
\newsavebox{\@brx}
\newcommand{\llangle}[1][]{\savebox{\@brx}{\(\m@th{#1\langle}\)}%
  \mathopen{\copy\@brx\kern-0.5\wd\@brx\usebox{\@brx}}}
\newcommand{\rrangle}[1][]{\savebox{\@brx}{\(\m@th{#1\rangle}\)}%
  \mathclose{\copy\@brx\kern-0.5\wd\@brx\usebox{\@brx}}}

\usepackage{graphicx}% Include figure files
\usepackage{dcolumn}% Align table columns on decimal point
\usepackage{bm}% bold math
\usepackage{xcolor}
\usepackage{xspace}
\usepackage[colorlinks,allcolors=blue]{hyperref}
\usepackage[makeroom]{cancel}
\usepackage{float}
\usepackage{siunitx}
\usepackage{color}
\definecolor{dgreen}{rgb}{0,0.7,0}
\def\bluew#1{{\color{red} #1}}

\def\bluew#1{{\color{black} #1}}
\newcommand{\beq}{\begin{equation}}
\newcommand{\eeq}{\end{equation}}
\newcommand{\bea}{\begin{eqnarray}}
\newcommand{\eea}{\end{eqnarray}}
\usepackage{subfigure}

\usepackage{natbib}
\usepackage[normalem]{ulem}

\begin{document}
\title{Inferring entropy production from time-dependent moments}

%\author{Prashant Singh$^1$ and Karel Proesmans}
%\email{prashant.singh@nbi.ku.dk}
%\email{karel.proesmans@nbi.ku.dk}

\author{Prashant Singh}
\email{prashant.singh@nbi.ku.dk}
\author{Karel Proesmans}
\email{karel.proesmans@nbi.ku.dk}

%\address{International Centre for Theoretical Sciences, Tata Institute of Fundamental ~Research, Bengaluru 560089, India $^1$}
\address{Niels Bohr International Academy, Niels Bohr Institute,
University of Copenhagen, Blegdamsvej 17, 2100 Copenhagen, Denmark}
%\ead{karel.proesmans@nbi.ku.dk}

%\ead{anupam.kundu@icts.res.in}
\vspace{10pt}

\begin{abstract}
Measuring entropy production of a system directly from the experimental data is highly desirable since it gives a quantifiable measure of the time-irreversibility for non-equilibrium systems and can be used as a cost function to optimize the performance of the system. Although numerous methods are available to infer the entropy production of stationary systems, there are only a limited number of methods that have been proposed for time-dependent systems and, to the best of our knowledge, none of these methods have been applied to experimental systems. Herein, we develop a general non-invasive methodology to infer a lower bound on the mean total entropy production for arbitrary time-dependent continuous-state Markov systems in terms of the moments of the underlying state variables. The method gives quite accurate estimates for the entropy production, both for theoretical toy models and for experimental bit erasure, even with a very limited amount of experimental data.
\end{abstract}
\maketitle

\section*{Introduction}
%\label{sec:intro}
In the last two decades, the framework of stochastic thermodynamics has enabled us to give thermodynamic interpretation to mesoscopic systems that are driven arbitrarily far from equilibrium \citep{Seifert-review}. Using the assumption of local detailed balance and a time scale separation between the system and its environment, one can define thermodynamic quantities such as heat, work and entropy production for general Markov systems \citep{Peliti2021, Sekimoto1998}. General results such as fluctuation theorems and thermodynamic uncertainty relations have been proposed and verified experimentally \citep{Seifert2005, Hayashi2010, TUR-1, TUR-2, TUR-3, TUR-4, TUR-5, TUR-6, TUR-7, TUR-8, TUR-9}. A fundamental quantity in stochastic thermodynamics is the entropy production \citep{Seifert-review, Peliti2021}. Indeed, minimising entropy production is an important problem in several applications, such as heat engines \citep{Pietzonka2018}, information processing \citep{Proesmans2020} and biological systems \citep{Ilker2022, MuruganKPR}. Furthermore, it provides information and places bounds on the dynamics of the system in the form of the thermodynamic speed limits \citep{TSL-1, TSL-2, TSL-3, TSL-4, TSL-5, TSL-6, TSL-7, TSL-8, TSL-9} and the dissipation-time uncertainty relation \citep{DTUR-1,DTUR-0, DTUR-2, DTUR-3}. Meanwhile measuring entropy production experimentally is difficult since one needs extensive knowledge about the details of the system, requiring a large amount of data. 

Over the last couple of years, several methods have been developed to infer entropy production and other thermodynamic quantities in a model-free way or with limited information about the underlying system \citep{Seifert-review-2, Roldanvp}. These techniques involve estimating entropy production using some experimentally accessible quantities. This includes methods based on thermodynamic bounds, such as the thermodynamic uncertainty relation \citep{Estimate-0,Estimate-1, Estimate-2}, inference schemes based on the measured waiting-time distribution \citep{WTT-1, WTT-2, WTT-3, WTT-4, WTT-5}, and methods based on ignoring higher-order interactions \citep{interactions-1, interactions-2}. However, most of these methods focus only on steady-state systems and do not work for time-dependent processes \citep{others-1, others-2, others-3, others-4, others-5, Baiesi-2023, Dechant2023new, new-1, new-2, new-3, new-4, new-5, new-6}. Nonetheless, time-dependent dynamics plays a crucial role in many systems. One can for example think about nano-electronic processes, such as bit erasure or AC-driven circuits, but also in biological systems, where oscillations in, e.g., metabolic rates or transcription factors, effectively lead to time-dependent driving \citep{Bit-1,Freitas2020,Glycolytic,Transcription}. 
There are few methods to infer entropy production for time-dependent systems \citep{TUR-6, Koyuk2020, TUR-7, Dechant2023, Otsubo2022, Lee2023}, and it is often challenging to apply them to experimental data. For instance, thermodynamic bound-based methods require the measurement of quantities such as response functions \citep{TUR-6, Koyuk2020} or time-inverted dynamics \citep{TUR-7}, which require an invasive treatment of the system. Another interesting method combines machine learning with a variational approach \citep{Otsubo2022}. This method does however require a large amount of experimental data and a high temporal resolution. 
This might also be the reason why, to the best of our knowledge, no experimental studies exist on the inference of entropy production under time-dependent driving. 

This paper aims to address these issues by deriving a general lower bound on the mean of the total entropy produced for continuous-state time-dependent Markov systems in terms of the time-dependent moments of the underlying state variables. The method gives an analytic lower bound for the entropy production if one only focuses on the first two moments and reduces to a numerical scheme if higher moments are taken into account.
The scheme also gives the optimal force field that minimizes the entropy production for the given time-dependent moments. After presenting the theoretical calculations and applications to analytically solvable models, we will test our method in an experimental scenario involving the erasure of computational bits \citep{Bit-1}. The method works quite well for all examples, even when only the first two moments are taken into account and with a very limited amount of data ($<100$ trajectories).

%Thus, not only the mean entropy production, we also get an estimate of the full probability distribution of the entropy production. In this way, we develop a method that allows us to infer the statistics of entropy production for arbitrary time-dependent systems using experimentally accessible quantities. We verify our method by applying it to the experimental data for bit erasure.  

%In the following, we first recall some known results from stochastic thermodynamics in Section \ref{sec-st}, which will then be used to derive bounds first for one-dimensional systems in Section \ref{sec-EP} and subsequently in general $d$-dimensional systems in Section \ref{gend-sec}. Subsequently, we  apply our method to the theoretical toy examples in Section \ref{application-sec} and to the experimental trajectories associated with erasure operation of computational bits in Section \ref{sec-BE}. Finally, we conclude in Section \ref{con-sec}.

\section*{Methods}
%\section*{Stochastic thermodynamics}
%\label{sec-st}
\noindent
\textbf{Stochastic thermodynamics.} We consider a $d$-dimensional stochastic system whose state at time $t$ is denoted by a variable $\bold{x}(t) = \{ x_1(t), x_2(t), ....,x_d(t) \}$. The system experiences a time-dependent force field $\bold{F} \left( \bold{x}(t),t \right)$ and is in contact with a thermal reservoir at constant temperature $T$. The stochastic variable $\bold{x}(t)$ evolves according to an overdamped Langevin equation
%\begin{align}
% \frac{d \bold{x}}{dt} = -\gamma^{-1}\bold{F} \left( \bold{x}(t),t \right) + \boldsymbol{\zeta}(t), 
%\label{Langevin}
%\end{align}
\begin{align}
 \frac{d \bold{x}}{dt} = \left( k_B T\right)^{-1}\bold{D}~\bold{F} \left( \bold{x}(t),t \right) + \boldsymbol{\zeta}(t), 
\label{Langevin}
\end{align}
$k_B$ is the Boltzmann constant and $\boldsymbol{\zeta}(t)$ represents the Gaussian white noise with $\left\langle\boldsymbol{\zeta}(t)\right\rangle=0$ and $\left\langle \boldsymbol{\zeta}(t)\boldsymbol{\zeta}(t')\right\rangle=2\bold{D}~\delta(t-t')$, with $\bold{D}$ being the diffusion matrix. One can also write the Fokker-Planck equation associated with the probability distribution $P \left( \bold{x},t \right)$ as
\begin{align}
\frac{\partial  P \left( \bold{x},t \right) }{\partial t} = -\bold{\nabla} \cdot \left[ \bold{v}\left( \bold{x},t \right)  P \left( \bold{x},t \right) \right], \label{Fokker-Planck}
\end{align} 
with $\bold{v}\left( \bold{x},t \right) $ being the probability flux given by
\begin{align}
\bold{v}\left( \bold{x},t \right)  = \frac{\bold{D}}{k_B T} 
\left[ \bold{F} \left( \bold{x},t \right)-k_B T~ \boldsymbol{\nabla} \log  P \left( \bold{x},t \right) \right]. \label{force-eqqvd}
\end{align}
This Fokker-Planck equation arises in a broad class of physical systems like colloidal particles \citep{Blickle2006}, electronic circuits \citep{Freitas2020} and spin systems \citep{Garanin1997}. Stochastic thermodynamics dictates that the mean total entropy produced when the system runs up to time $t_f$ is given by \citep{Peliti2021}
\begin{align}
S_{\text{tot}}(t_f) =k_B\int_{0}^{t_f} dt ~\int _{-\infty}^{\infty}d \bold{x} ~P \left( \bold{x},t \right) \bold{v}\left( \bold{x},t \right) \bold{D}^{-1}\bold{v}\left( \bold{x},t \right) . \label{entropy-production}
\end{align}
Although this formula, in principle, allows one to calculate the entropy production of any system described by a Fokker-Planck equation, one would need the force field and probability distribution at each point in space and time. This is generally not feasible in experimental set-ups.
%This equation clearly suggests that one needs to know both force-field $\bold{F} \left( \bold{x},t \right)$ and distribution function $P \left( \bold{x},t \right)$ to calculate $S_{\text{tot}}(t_f) $. However this poses two practical difficulties in real physical systems. First is - experimentally one might not always have a detailed knowledge of the underlying force field $\bold{F} \left( \bold{x},t \right)$. Second is - even if one knows $\bold{F} \left( \bold{x},t \right)$, one still needs a large number of samples to obtain $P \left( \bold{x},t \right)$. While this can be done in simulations, it might not always be possible in experiments to generate such large sample sizes. 

In this paper, our aim is to develop a technique that gives information about the total entropy production in terms of experimentally accessible quantities. Using methods from the optimal transport theory \citep{Villani2003, Benamou2000, TSL-1, KSaito2023}, we will derive a non-trivial lower bound on $S_{\text{tot}}(t_f)$ completely in terms of the moments of $\bold{x}(t)$. This means, we can infer an estimate for $S_{\text{tot}}(t_f)$ simply by measuring the first few moments. \\
%Our method also gives the complete form of the optimal flux $\bold{v}\left( \bold{x},t \right)$  that saturates the lower bound.

%\section*{Bound for one-dimensional systems}
%\label{sec-EP}

\noindent
\textbf{Bound for one-dimensional systems.} Let us first look at the one-dimensional case $(d=1)$ and denote the $n$-th moment of $x(t)$ by $X_n(t)=\langle x(t)^n\rangle$. Throughout this paper, we assume that these moments are finite for the entire duration of the protocol. We will first look at the case where one only has access to the first two moments $X_1(t)$ and $X_2(t)$, as the resulting estimate for the entropy production has a rather simple and elegant expression. Subsequently we will turn to the more general case. \\
%\subsection{First two moments}
%\label{sec-mom2}

\noindent
\textit{First two moments:} Our central goal is to obtain a lower bound on $S_{\text{tot}}(t_f)$, constrained on the known time-dependent moments. This can be done by minimizing the following action:
%\begin{align}
%\mathbb{S}\left( x,v, t_f \right) &= \frac{D}{k_B}  S_{\text{tot}}(t_f) + \int _0 ^{t_f} dt~ \left[ \mu _1(t) X_1(t)+\mu _2(t) X_2(t) \right]+ \zeta _1(t_f) X_1(t_f)+\zeta _2(t_f) X_2(t_f), \nonumber \\
%& = \int _0 ^{t_f} dt \int _{-\infty}^{\infty}dx ~P(x,t) \left[ v^2(x,t)+ \mu _1(t) x + \mu _2 (t) x^2   \right] + \int _{-\infty}^{\infty}dx ~P(x,t_f) \left[ \zeta _1(t_f) x+\zeta _2(t_f) x^2\right],  \label{action-eq-1}
%\end{align}
\begin{widetext}
\begin{align}
\mathbb{S}\left( x,v, t_f \right) &= \frac{D}{k_B}  S_{\text{tot}}(t_f) + \int _0 ^{t_f} dt~ \left[ \mu _1(t) X_1(t)+\mu _2(t) X_2(t) \right]+ \zeta _1(t_f) X_1(t_f)+\zeta _2(t_f) X_2(t_f), \nonumber \\
& = \int _0 ^{t_f} dt \int _{-\infty}^{\infty}dx ~P(x,t) \left[ v^2(x,t)+ \mu _1(t) x + \mu _2 (t) x^2   \right] + \int _{-\infty}^{\infty}dx ~P(x,t_f) \left[ \zeta _1(t_f) x+\zeta _2(t_f) x^2\right],  \label{action-eq-1}
\end{align}
\end{widetext}
with respect to the driving protocol, $F(x,t)$. We have introduced four Lagrange multipliers, two of which, namely $\zeta _1(t_f)$ and $\zeta _2(t_f)$ associated with fixing of first two moments at the final time $t_f$ and the other two, $\mu _1(t)$ and $\mu _2(t)$ at the  intermediate times. In the action, we have included the contribution of moments at the final time $t_f$ separately, as dictated by the Pontryagin’s Maximum Principle \citep{Pontryagin}
%. This turns out to be rather crucial for deriving appropriate boundary condition for the optimal path as we discuss later.} 

\noindent
Minimizing the action with respect to $F(x,t)$ seems very complicated, due to the non-trivial dependence of $P(x,t)$ on $F(x,t)$ in Eq.~\eqref{Fokker-Planck}. 
%In fact, we show in \ref{appen-rate} that minimising with respect to $F(x,t)$ directly leads to sub-optimal protocols. \textbf{\redw{(Discuss with Karel)}} 
To circumvent this problem, we use methods from the optimal transport theory and introduce a coordinate transformation $y(x,t)$ as \citep{Benamou2000}
\begin{align}
\dot{y}(x,t) = v\left( y(x,t),t \right),~~~~~\text{with }y(x,0) = x,
\label{Lag-cord-eq-1}
\end{align}
and overdot indicating the time derivative. Mathematically, this coordinate transformation is equivalent to changing from Eulerian to Lagrangian description in fluid mechanics. It has also been used to derive other thermodynamic bounds such as the thermodynamic speed limit \citep{TSL-1}. Using this equivalence, one can show that \citep{Batchelor2000}
\small{\begin{align}
 \int _{-\infty}^{\infty}dx~ P(x,t) ~g(x,t)  &= \int _{-\infty}^{\infty}dx ~P_0(x)~ g(y(x,t),t), \label{Lag-cord-eq-22} \\
\int _{-\infty}^{\infty}dx P(x,t) v(x,t)g(x,t)  &= \int _{-\infty}^{\infty}dx P_0(x) \frac{\partial y(x,t)}{\partial t}g(y(x,t),t),
\label{Lag-cord-eq-2}
\end{align}}
\normalsize{for} any test function $g(x,t)$ and initial probability distribution $P_0(x) = P(x,0)$. Plugging $g(x,t) = \delta \left(x- y(x',t)\right)$, we get
\begin{align}
P\left( y(x',t),t \right) = \frac{P_0(x')}{|\partial _{x'} y(x',t)|}. \label{OTT-dist}
\end{align}
On the other hand, putting $g(x,t) = v(x,t)$ in Eq.~\eqref{Lag-cord-eq-2}, we obtain
\begin{align}
S_{\text{tot}}(t_f) =\frac{k_B}{D}  \int_{0}^{t_f} dt ~\int _{-\infty}^{\infty}dx ~P_0 \left( x\right) ~\dot{y}\left( x,t \right) ^2. \label{ED-erfs}
\end{align}
Similarly, the moments and their time-derivatives are given by
\begin{align}
X_n(t) &=   \int _{-\infty}^{\infty}dx~y(x,t)^n~ P_0(x),  \label{mom-M2-eq-1} \\
\bluew{\dot{X}_n(t) }&  =  \bluew{n \int _{-\infty}^{\infty}dx ~y(x,t)^{n-1}~\dot{y}(x,t)~P_0(x).} \label{der-mom-M2-eq-1}
\end{align}
%\begin{align}
%X_n(t) &= \int _{-\infty}^{\infty}dx~x^n~ P(x,t) =   \int _{-\infty}^{\infty}dx~y(x,t)^n~ P_0(x),  \label{mom-M2-eq-1} \\
%\dot{X}_n(t) &  =  n \int _{-\infty}^{\infty}dx ~x^{n-1}~P(x,t)~v(x,t) = n \int _{-\infty}^{\infty}dx ~y(x,t)^{n-1}~P(x,t)~\dot{y}(x,t). \label{der-mom-M2-eq-1}
%\end{align}
We can also write the force-field $F(x,t)$ in terms of $y(x,t)$ by using Eqs.~\eqref{force-eqqvd} and \eqref{Lag-cord-eq-1}. Hence we can calculate various quantities such as probability distribution, mean total dissipation, and force field utilizing $y(x,t)$. In fact, it turns out that $y(x,t)$ is uniquely determined for a given force-field. Using the relations in Eqs.~\eqref{Lag-cord-eq-22} and \eqref{Lag-cord-eq-2}, we can reformulate the optimisation of the action $\mathbb{S}\left( x,v, t_f \right)$ in Eq.~\eqref{action-eq-1} with respect to $F(x,t)$ as an optimisation problem in $y(x,t)$. Meanwhile one can rewrite $\mathbb{S}\left( x,v, t_f \right)$ as
%\begin{align}
%\mathbb{S}\left(y, \dot{y}, t_f \right) =& \int _0 ^{t_f} dt \int _{-\infty}^{\infty}dx P_0(x) \left[ \dot{y}(x,t)^2 + \mu _1(t) y(x,t) + \mu _2(t) y(x,t)^2 \right] + \int _{-\infty}^{\infty}dx P_0(x) \left[  \zeta _1(t_f) y(x,t_f) + \zeta _2(t_f) y(x,t_f)^2 \right].
%\label{action-eq-2}
%\end{align}
%For a small change $y(x,t) + \delta y(x,t)$, the total change in action is
%\begin{align}
%\delta \mathbb{S} = &  \int _0 ^{t_f} dt \int _{-\infty}^{\infty}dx~ P_0(x) ~\left[ -2 \ddot{y}(x,t) + \mu _1(t)+2 \mu _2(t) y(x,t)\right] ~\delta y(x,t) - 2\int _{-\infty}^{\infty}dx~ P_0(x) ~ \dot{y}(x,0) ~ \delta y(x,0).\nonumber \\
%& ~~~~~~~+ \int _{-\infty}^{\infty}dx~ P_0(x) ~\left[ 2 \dot{y}(x,t_f)+ \zeta _1(t_f)+2 \zeta _2(t_f) y(x,t_f)\right] ~\delta y(x,t_f).
%\label{action-eq-3}
%\end{align}
\begin{widetext}
\begin{align}
\mathbb{S}\left(y, \dot{y}, t_f \right) =& \int _0 ^{t_f} dt \int _{-\infty}^{\infty}dx P_0(x) \left[ \dot{y}(x,t)^2 + \mu _1(t) y(x,t) + \mu _2(t) y(x,t)^2 \right] + \int _{-\infty}^{\infty}dx P_0(x) \left[  \zeta _1(t_f) y(x,t_f) + \zeta _2(t_f) y(x,t_f)^2 \right].
\label{action-eq-2}
\end{align}
For a small change $y(x,t) + \delta y(x,t)$, the total change in action is
\begin{align}
\delta \mathbb{S} = &  \int _0 ^{t_f} dt \int _{-\infty}^{\infty}dx~ P_0(x) ~\left[ -2 \ddot{y}(x,t) + \mu _1(t)+2 \mu _2(t) y(x,t)\right] ~\delta y(x,t) - 2\int _{-\infty}^{\infty}dx~ P_0(x) ~ \dot{y}(x,0) ~ \delta y(x,0).\nonumber \\
& ~~~~~~~+ \int _{-\infty}^{\infty}dx~ P_0(x) ~\left[ 2 \dot{y}(x,t_f)+ \zeta _1(t_f)+2 \zeta _2(t_f) y(x,t_f)\right] ~\delta y(x,t_f).
\label{action-eq-3}
\end{align}
\end{widetext}
For the optimal protocol, this small change has to vanish. Vanishing of the first term on the right hand side for arbitrary $\delta y(x,t)$ gives the Euler-Lagrangian equation
\begin{align}
2 \ddot{y}(x,t) = \mu _1(t) + 2 \mu _2(t) y(x,t). \label{euler-eq-1}
\end{align}
Let us now analyse the other two terms. Recall that for all possible paths $y(x,t)$, optimisation is being performed with the fixed initial condition $y(x,0) = x$ [see Eq.~\eqref{Lag-cord-eq-1}]. Consequently, we have $\delta y(x,0) = 0$ which ensures that the second term on the right hand side of Eq.~\eqref{action-eq-3} goes to zero. On the other hand, the value of $y(x,t_f)$ at the final time can be different for different paths which gives $\delta y(x,t_f) \neq 0 $. For the second line to vanish, we must then have the pre-factor associated with $\delta y(x,t_f) $ equal to zero. Thus, we get two boundary conditions in time
\begin{align}
2 \dot{y}(x,t_f) &= - \zeta _1(t_f) - 2 \zeta _2(t_f) y(x,t_f), \label{euler-eq-2} \\
 y(x,0) & = x. \label{euler-eq-3} 
\end{align}
Observe that the first equation clearly demonstrates the importance of incorporating the terms at the final time $t_f$ in the action described in Eq.~\eqref{action-eq-1}. If these terms are omitted, we would obtain $\dot{y}(x,t_f) = 0$, which according to Eq.~\eqref{der-mom-M2-eq-1} suggests that the time derivative of moments at the final time is zero. However, this is not true. To get consistent boundary condition, it is necessary for $\zeta_1(t_f)$ and $\zeta_2(t_f)$ to be non-zero.

\noindent
Proceeding to solve Eq.~\eqref{euler-eq-1}, one can verify that
\begin{align}
\dot{y}(x,t) = \lambda _1(t)+ \lambda _2(t) ~y(x,t), \label{y-genn-eq-1}
\end{align}
where $\lambda_1(t)$ and $\lambda_2(t)$ are functions that are related to the Lagrange multipliers $\mu _1(t)$ and $\mu _2(t)$. To see this relation, one can take the time derivative in Eq.~\eqref{y-genn-eq-1} and compare it with the Euler-Lagrange equation \eqref{euler-eq-1} to obtain
\begin{align}
\begin{cases}
\mu _1(t) &= 2 \left( \dot{\lambda}_1(t) +\lambda _1(t) \lambda _2(t) \right), \\
\mu _2(t) & = \dot{\lambda}_2(t) + \lambda _2(t)^2. 
\end{cases}
\end{align}
%\begin{align}
%\mu _1(t) = 2 \left( \dot{\lambda}_1(t) +\lambda _1(t) \lambda _2(t) \right)~~\text{and}~~\mu _2(t) = \dot{\lambda}_2(t) + \lambda _2(t)^2. 
%\end{align}
Let us now compute these $\lambda(t)$-functions. For this, we use the solution of $\dot{y}(x,t) $ in $\dot{X}_n(t)$ in Eq.~\eqref{der-mom-M2-eq-1} for $n=1$ and $n=2$. This yields
\begin{align}
\begin{cases}
\lambda _1(t) &= \frac{2 X_2(t) \dot{X}_1(t)-X_1(t) \dot{X}_2(t)}{2(X_2(t)-X_1^2(t))}, \\
\lambda _2(t) & = \frac{\dot{X}_2(t)-2 X_1(t) \dot{X}_1(t)}{2(X_2(t)-X_1^2(t))}. 
\end{cases}
\label{eq-v-lam}
\end{align}
%\begin{align}
%\lambda _1(t) = \frac{2 X_2(t) \dot{X}_1(t)-X_1(t) \dot{X}_2(t)}{2(X_2(t)-X_1^2(t))}~~\text{and}~~\lambda _2(t) = \frac{\dot{X}_2(t)-2 X_1(t) \dot{X}_1(t)}{2(X_2(t)-X_1^2(t))}. \label{eq-v-lam}
%\end{align}
%\begin{align}
%a _1(t) & = \frac{X_1(t)\dot{X}_2(t)-2 X_2(t)\dot{X}_1(t)}{X_2(t)- X_1(t)^2}, \\
%a _2 (t)& = \frac{2X_1(t)\dot{X}_1(t)-\dot{X}_2(t)}{2 \left[  X_2(t)- X_1(t)^2 \right]}.
%\end{align}
One can then use Eq.~\eqref{ED-erfs} to show that $S_{\text{tot}}(t_f) $ satisfies the bound
\begin{align}
S_{\text{tot}}(t_f) \geq ~S_{\text{tot}}^{\text{12}}(t_f) = \frac{k_B}{D}\int_0^{t_f} dt ~\left[     \frac{\dot{A}_2(t)^2}{4 A _2(t) }  + \dot{X}_1(t)^2  \right], \label{estimate-eq-1}
\end{align}
where $A _2 (t) = X_2(t)-X_1(t)^2$ is the variance of $x(t)$ and $S_{\text{tot}}^{\text{12}}(t_f)$ is the bound on the mean total entropy produced with first and second moments fixed. Eq.~\eqref{estimate-eq-1} is one of the central results of our paper. It gives a lower bound $S_{\text{tot}}^{\text{12}}(t_f)$ on the mean total entropy dissipated completely in terms of the first moment and variance. \bluew{Our bound also requires the knowledge of the diffusion coefficient. For this, one can use the short-time mean squared displacement
\begin{align}
D =  \lim _{\delta t \to 0}\frac{\langle \delta x^2(t)  \rangle}{2 ~\delta t},~~\text{where }\delta x = x(t+\delta t)-x(t),
\end{align}
as used recently in \citep{Lee2023} or use other inferring methods \citep{Dean2012, Gerlee2023}. In this work, we do not delve into the estimation of diffusion coefficient and assume it to be given.}

%It is quite natural to ask for which process is this bound saturated. We show in  \ref{appen-harmonic} that this is the case for Gaussian processes.

%Notice that, throughout our analysis, we have minimised the average total entropy produced till time $t_f$ with respect to the force field. However, one can also minimise the entropy production rate (instead of the total entropy) at all intermediate times. As shown in \ref{appen-rate}, this generally leads to sub-optimal protocols.

Furthermore, in our analysis, we have focused on minimizing the mean total entropy generated up to time $t_f$ with respect to the force field. However, an alternative approach involves minimizing the entropy production rate at each intermediate time, as opposed to the total entropy production, but, in Supplementary note 1, we show that this approach generally leads to sub-optimal protocols and therefore no longer gives a bound on the total entropy production.\\

%\subsection{Optimal protocol with first two moments}
\noindent
\textit{Optimal protocol with first two moments:} Having obtained a lower bound for the entropy production, it is a natural question to ask for which protocols this bound is saturated. To answer that question, we first turn to the flux $v(x,t)$ that gives rise to this optimal value. For this, we use Eqs.~\eqref{Lag-cord-eq-1} and \eqref{y-genn-eq-1} and obtain
\begin{align}
v(x,t) =v^{\text{12}}(x,t) = \lambda _1(t) + \lambda _2(t) x, \label{eq-v-fm}
\end{align}
with $\lambda(t)$-functions given in Eq.~\eqref{eq-v-lam}. One can verify that this equation is always satisfied for Gaussian processes. This is shown in Supplementary note 2. 

\noindent
Next, we turn to the probability distribution associated with the optimal dissipation. For this, one needs the form of $y(x,t)$ which follows from Eq.~\eqref{y-genn-eq-1}
\begin{align}
& y(x,t)  = \alpha(t) x  + \beta (t), \label{y-genn-eq-2} \\
\text{with }&\alpha(t) = e^{  \int _0 ^{t}dt' \lambda _2(t')},~\beta(t) = \alpha(t) \int _0 ^{t}dt'~\frac{\lambda _1(t')}{\alpha(t')}.
\end{align}
Plugging this in Eq.~\eqref{OTT-dist}, we obtain the distribution as
\begin{align}
P(x,t) = P^{\text{12}}(x,t)=\frac{1}{\alpha(t)} P_0 \left( \frac{x-\beta(t)}{\alpha(t)}\right), \label{eqq-prob-ps1}
\end{align}
which then gives the optimal force-field as
\begin{equation}
\scalebox{0.91}{$
\begin{split}
F^{\text{12}}\left( x,t \right)  = \frac{k_B T}{D} \left[  \lambda _1(t) + \lambda _2(t)~x + D \frac{\partial }{\partial x} \ln P_0 \left( \frac{x-\beta(t)}{\alpha(t)}\right) \right].
\end{split}$}
\end{equation}
We see that the optimal protocol comprises solely of a conservative force field with the associated energy landscape characterized by two components: the first two terms correspond to a time-dependent harmonic oscillator, whereas the last term has the same shape as the equilibrium landscape associated with the initial state. Furthermore, from Eq.~\eqref{eqq-prob-ps1}, we also observe that this optimal protocol preserves the shape of the initial probability distribution. A similar result was also recently observed in the context of the precision-dissipation trade-off relation for driven stochastic systems \citep{Proesmans2023}.\\

%This signifies that the optimal protocol consists only of a conservative force-field with energy landscape that has two contributions- one is the time-dependent harmonic oscillator and the other is a landscape that depends on the initial energy landscape. Following Eq. \eqref{eqq-prob-ps1}, we also find that the optimal protocol conserves the overall shape of the initial probability distribution.
%gradient-type force field signifies that the ideal protocol exclusively relies on a conservative energy landscape, and introducing non-conservative forces typically does not enhance the accuracy of the driving process, but instead introduces additional dissipation.
%With this force-field, we can now simulate the Langevin equation \eqref{Langevin} and obtain the full probability distribution of total dissipated entropy.
%\subsection{Fixing first $m$ moments}
%\label{d-1-genm-sec}
\noindent
\textit{Fixing first $m$ moments:} So far, we have derived a bound on $S_{\text{tot}}(t_f) $ based only on the knowledge of first two moments. We will now generalise this for arbitrary number of moments. Consider the general situation where the first $m$ moments of the variable $x(t)$ are given. Our goal is to optimise the action
\begin{equation}
\scalebox{0.89}{$
\begin{split}
\mathbb{S}\left( x,v, t_f\right) & = \frac{D}{k_B} S_{\text{tot}}(t_f)+  \sum _{i = 1}^{m} \Big[ \int _0 ^{t_f} dt~\mu _i(t)~X_i(t)  +  \zeta _i(t_f) X_i(t_f) \Big],  \\
& = \int _0 ^{t_f} dt \int _{-\infty}^{\infty}dx ~P(x,t)~ \left[ v^2(x,t)+ \sum _{i=1}^{m}\mu _i(t) ~x^i \right]   \\
&~~~~~~~ +\sum _{i=1}^{m} \int _{-\infty}^{\infty}dx~ P(x,t_f) ~ \zeta _i(t) ~x^i,
\end{split}$}
\label{action-gen-M2-eq-1}
\end{equation}
with respect to $v(x,t)$. Once again, $\mu(t)$- functions are the Lagrange multipliers for $m$ moments with contribution at the final time $t= t_f$ included separately. As done before in Eq.~\eqref{Lag-cord-eq-1}, we introduce the new coordinate $y(x,t)$, and rewrite  the action in terms of this map
\begin{equation}
\scalebox{0.91}{$
\begin{split}
\mathbb{S}\left(y, \dot{y}, t_f \right) =& \int _0 ^{t_f} dt \int _{-\infty}^{\infty}dx ~P_0(x)~ \left[ \dot{y}(x,t)^2 + \sum _{i=1}^{m}\mu _i(t) ~y(x,t)^i \right] \nonumber \\
&~~~~+ \int _{-\infty}^{\infty}dx~ P_0(x)  ~\sum _{i=1}^{m}\zeta _i(t_f) ~y(x,t_f)^i.
\end{split}$}
\nonumber
\end{equation}
For a small change $y(x,t) + \delta y(x,t)$, the total change in action is
%\small{\begin{align}
%\delta \mathbb{S} = & \int _0 ^{t_f} dt \int _{-\infty}^{\infty}dx ~P_0(x)~ \left[-2 \ddot{y}(x,t) 
 %+ \sum _{i=1}^{m}i \mu _i(t) ~y(x,t)^{i-1} \right]~\delta y(x,t) -2  \int _{-\infty}^{\infty}dx~ P_0(x)  \dot{y}(x, 0)~\delta y(x,0) \nonumber\\
 %& + \int _{-\infty}^{\infty}dx~ P_0(x)  \left[2 \dot{y}(x, t_f)+\sum _{i=1}^{m}i ~\zeta _i(t_f) ~y(x,t_f)^{i-1} \right]~\delta y(x,t_f),
%\end{align}}
\begin{widetext}
\begin{align}
\delta \mathbb{S} = & \int _0 ^{t_f} dt \int _{-\infty}^{\infty}dx ~P_0(x)~ \left[-2 \ddot{y}(x,t) 
 + \sum _{i=1}^{m}i \mu _i(t) ~y(x,t)^{i-1} \right]~\delta y(x,t) -2  \int _{-\infty}^{\infty}dx~ P_0(x)  \dot{y}(x, 0)~\delta y(x,0) \nonumber\\
 &~~~~~ + \int _{-\infty}^{\infty}dx~ P_0(x)  \left[2 \dot{y}(x, t_f)+\sum _{i=1}^{m}i ~\zeta _i(t_f) ~y(x,t_f)^{i-1} \right]~\delta y(x,t_f),
\end{align}
\end{widetext}
%\small{\begin{align}
%\delta \mathbb{S} = & \int _0 ^{t_f} dt \int _{-\infty}^{\infty}dx ~P_0(x)~ \left[-2 \ddot{y}(x,t) 
 %+ \sum _{i=1}^{m}i \mu _i(t) ~y(x,t)^{i-1} \right]~\delta y(x,t) \nonumber\\
 %& + \int _{-\infty}^{\infty}dx~ P_0(x)  \left[2 \dot{y}(x, t_f)+\sum _{i=1}^{m}i ~\zeta _i(t_f) ~y(x,t_f)^{i-1} \right]~\delta y(x,t_f) \nonumber \\
%& ~~~~~~ -2  \int _{-\infty}^{\infty}dx~ P_0(x)  \dot{y}(x, 0)~\delta y(x,0),
%\end{align}}
\normalsize{which} vanishes for the optimal path. Following the same line of reasoning as before, we then obtain the Euler-Lagrange equation
\begin{align}
2 \ddot{y}(x,t) = \sum _{i=1}^{m}i~ \mu _i(t) ~y(x,t)^{i-1}, \label{euler-gen-eq-1}
\end{align}
with two boundary conditions
\begin{align}
2 \dot{y}(x, t_f) & = -\sum _{i=1}^{m}i~ \zeta _i(t_f) ~y(x,t_f)^{i-1},  \label{euler-gen-eq-2}  \\
y(x,0) & = x. \label{euler-gen-eq-3} 
\end{align}
However, we still need to compute functions $\mu_i(t)$ and $\zeta_i(t)$ with $i=1,2,..,m$ in order to completely specify the boundary conditions. To calculate $\zeta$-functions, we use Eq.~\eqref{der-mom-M2-eq-1} and obtain
\begin{align}
- \frac{2}{i} \dot{X}_i (t_f) & = \zeta _1(t_f) X_{i-1}(t_f) +2 \zeta _2(t_f) X_{i}(t_f) +... \nonumber \\
&~~~~+...+m \zeta _m(t_f) X_{m-2+i}(t_f). \label{gen-lin-eq-1}
\end{align}
On the other hand, for $\mu(t)$-functions, we use the second derivative of moments and obtain
\begin{align}
 \ddot{X}_i(t) &= \mathcal{B} _i(t) + \frac{i}{2} \left[ \mu _1(t) X_{i-1}(t)+2 \mu_2(t) X_{i}(t)+...\right.\nonumber \\
 &~~~~~\left.+...+m \mu _m(t) X_{m-2+i}(t) \right] , \label{gen-lin-eq-2}\\
 \text{with   }\mathcal{B}_i(t) &= i(i-1) \int _{-\infty}^{\infty} dx ~P_0(x) ~\dot{y}(x,t)^2~ y(x,t)^{i-2}.\label{gen-lin-eqdfa-2}
\end{align}
Both Eqs.~\eqref{gen-lin-eq-1} and \eqref{gen-lin-eq-2} hold for all positive integer values of $i \leq m$. Solving this gives all $\mu(t)$ and $\zeta(t)$ functions in terms of $\mathcal{B}_i$, $i=1,...,n$ and $X_i$, $i=1,...,m+n-2$.

Once Eq.~\eqref{euler-gen-eq-1} is solved, the bound on mean total entropy dissipated can then be written as
\begin{equation}
\scalebox{0.93}{$
\begin{split}
S_{\text{tot}}(t_f) \geq S_{\text{tot}}^{\text{12..m}}(t_f) = \frac{k_B}{D}\int _0^{t_f} dt \int _{-\infty}^{\infty}dx ~P_0(x)~\dot{y}(x,t)^2, 
\end{split}$}
\label{estimate-eq-2}
\end{equation}
\noindent
Unfortunately, we could not obtain an analytic expression for Eq.~\eqref{euler-gen-eq-1} due to the presence of higher-order moments and $\mathcal{B}_i$, which are not a priori known. Instead, we will focus on numerically integrating  Eq.~\eqref{euler-gen-eq-1}. One can use the value of $y(x,t)$ to calculate the higher-order moments and therefore the $\mu(t)$'s in Eq.~\eqref{gen-lin-eq-2}, which in turn can be used to obtain $y(x, t+\Delta t)$. Finally, one can make sure that the boundary condition, Eq.~\eqref{euler-gen-eq-2}, is satisfied through a shooting method.\\

%In comparison to $m \leq 2$, we observe an important difference when solving these equations for $m > 2$. For $m=2$, the right-hand sides involve only the first two moments and their time derivatives. Contrarily, for any $m>2$, they also involve higher moments. For instance, for $m=3$, we also need the knowledge of the fourth moment $X_4(t)$. This makes it impossible to derive an analytical expression for $y(x,t)$. Instead, we will focus on numerically integrating  Eq.~\eqref{euler-gen-eq-1}. One can use the value of $y(x,t)$ to calculate the higher-order moments and therefore the $\mu(t)$'s in Eq.~\eqref{gen-lin-eq-2}, which in turn can be used to obtain $y(x, t+\Delta t)$. Finally, one can make sure that the boundary condition, Eq.~\eqref{euler-gen-eq-2}, is satisfied through a shooting method.

%\section*{Bound in higher dimensions}
%\label{gend-sec}

\noindent
\textbf{Bound in higher dimensions.} Having developed a general methodology in one dimension, we now extend these ideas to higher dimensional systems with state variables $\bold{x}(t) = \{ x_1(t), x_2(t), ....,x_d(t) \}$. Herein again, our aim is to obtain a bound on $S_{\text{tot}}(t_f)$ with information about moments, including mixed moments.\\ 

%\subsection{First two moments}
\noindent
\textit{First two moments: }
As before, let us first introduce the notation 
%\begin{align}
%X_1^{(i)}(t)=\int _{-\infty}^{\infty}d \bold{x} ~x_i~P \left( \bold{x},t \right),~~~~~\mathcal{M}_{ij}(t)=\int _{-\infty}^{\infty}d \bold{x} ~x_i x_j~P \left( \bold{x},t \right),
%\end{align}
%\begin{align}
%& X_1^{i}(t)=\int _{-\infty}^{\infty}d \bold{x} ~x_i~P \left( \bold{x},t \right),\\
%& X_2^{i,j}(t)=\int _{-\infty}^{\infty}d \bold{x} ~x_i x_j~P \left( \bold{x},t \right), \label{bxgapw}
%\end{align}
\begin{equation}
\begin{split}
& X_1^{i}(t)=\int _{-\infty}^{\infty}d \bold{x} ~x_i~P \left( \bold{x},t \right),\\
& X_2^{i,j}(t)=\int _{-\infty}^{\infty}d \bold{x} ~x_i x_j~P \left( \bold{x},t \right),
\end{split} 
\label{bxgapw}
\end{equation}
for first two (mixed) moments of $x_i(t)$ variables. Once again, this is a constrained optimisation problem and we introduce the following action functional 
\footnotesize{\begin{align}
& \mathbb{S} \left( \bold{x}, \bold{v}, t_f\right)  =  \frac{D}{k_B} S_{\text{tot}}(t_f)+\int _0 ^{t_f} dt ~\left[  \sum _{i=1}^{d} \mu _i(t) X_1^{i}(t)  \right. \nonumber \\
& \left.+ \sum _{i,j=1}^{d} \lambda _{ij}(t) ~X_2^{i,j}(t)  \right] + \sum _{i=1}^{d} \alpha _i(t_f) X_1^{i}(t_f)   + \sum _{i,j=1}^{d} \beta _{ij}(t_f) ~X_2^{i,j}(t_f),  \\
& =\int _0 ^{t_f} dt \int _{-\infty}^{\infty} d \bold{x}~ P \left( \bold{x},t \right) \left[ \bold{v}\left( \bold{x},t \right) ^2 +    \sum _{i=1}^{d} \mu _i(t) ~x_i + \sum _{i,j=1}^{d} \lambda _{ij}(t) ~x_i x_j  \right] \nonumber \\
&  + \int _{-\infty}^{\infty} d \bold{x}~ P \left( \bold{x},t_f \right) \left[    \sum _{i=1}^{d} \alpha _i(t_f) ~x_i  + \sum _{i,j=1}^{d} \beta _{ij}(t_f) ~x_i x_j  \right], \label{genn-actty}
\end{align}}
\normalsize{where} functions $\mu_i(t)$, $\lambda_{ij}(t)$, $\alpha_i(t)$ and $\beta_{ij}(t)$ all stand for Lagrange multipliers associated with given constraints. Further carrying out this optimisation is challenging since distribution $P \left( \bold{x},t \right)$ depends non-trivially on the force-field $\bold{F} \left( \bold{x},t \right)$. We, therefore, introduce a $d$-dimensional coordinate
\begin{align}
\dot{\bold{y}}(\bold{x},t) = \bold{v} \left( \bold{y}(\bold{x},t),t\right),~~~~~\text{with  } \bold{y}(\bold{x},0) = \bold{x}, \label{genn-transport-map}
\end{align}
and rewrite $\mathbb{S}\left( \bold{x}, \bold{v}, t_f\right) $ completely in terms of $\bold{y}(\bold{x},t)$. We then follow same steps as done before but in higher dimensional setting.
Since this is a straightforward generalisation, we relegate this calculation to Supplementary note 3 and present only the final results here. For the optimal protocol, we find
\begin{align}
 \dot{y}_i\left( \bold{x} , t\right)= \dot{X}_1^{i}(t)+ \sum_{j=1}^{d} \frac{\gamma_{ij}(t)}{2} ~\left[ y_j\left( \bold{x} , t\right)  -X_1^{j}(t) \right], \label{high-dim-ohfakd}
\end{align}
with functions $\gamma_{ij}(t)$ related to the Lagrange multipliers $\mu_i(t)$, $\lambda_{ij}(t)$, see Eq.~(S34) in the Supplementary note. Furthermore, in Supplementary note 3, we show that $\gamma_{ij}(t)$ functions are the solutions of a set of linear equations:
%\begin{align}
%-2 \dot{A} _{ij}(t) &= \sum _{l=1}^{d} \left[ \tilde{\beta} _{il}(t) A_{jl}(t)  +   \tilde{\beta} _{jl}(t) A_{il}(t)   \right], \label{high-eqqdfah}
%\end{align}
\begin{align}
-2 \dot{\bold{A}}(t) = \boldsymbol{\gamma}(t) \bold{A}(t)+\bold{A}(t) \boldsymbol{\gamma}(t). \label{gen-eq1}
\end{align}
where $\boldsymbol{\gamma}(t)$ and $\bold{A}(t)$ are two matrices with elements $ \gamma _{ij}(t)$ and $A _{ij}(t) = X_2 ^{i,j}(t) - X _{1}^{i} (t)X _{1}^{j} (t) $ respectively. Notice that the elements of $\bold{A}(t)$ comprise only of variance and covariance which are known. Therefore, the matrix $\bold{A}(t)$ is fully known at all times. 

\noindent
To solve Eq.~\eqref{gen-eq1}, we note that $\bold{A}(t)$ is a symmetric matrix and thus, there exists an orthogonal matrix $\bold{O}(t)$ such that
\begin{align}
\bold{A}(t) =  \bold{O}(t)~\boldsymbol{\Lambda}(t)~\bold{O}^{T}(t), \label{gen-eq2}
\end{align}
where $\boldsymbol{\Lambda}(t)$ is a diagonal matrix whose elements correspond to the (positive) eigenvalues of $\bold{A}(t) $. $\bold{O}(t)$ can be constructed using eigenvectors of $\bold{A}(t)$ and it satisfies the orthogonality condition $\bold{O}^{-1}(t) =  \bold{O}^{T}(t)$. Combining the transformation Eq.~\eqref{gen-eq2} with Eq.~\eqref{gen-eq1}, we obtain
\begin{align}
{\mathcal{G}}_{ij}(t) &= -\frac{2 ~\left[   \bold{O}^{T}(t)~\dot{\bold{A}}~ \bold{O}(t) \right]_{ij}}{ {\Lambda}_{ii}(t)+{\Lambda}_{jj}(t)  } , \label{gen-eq3} \\ 
\text{with }\boldsymbol{\mathcal{G}}(t) &  =  \bold{O}^{T}(t)~\boldsymbol{\gamma}(t)~ \bold{O}(t). \nonumber 
\end{align}
This relation enables us to write $ \dot{y}_i\left( \bold{x} , t\right)$ in Eq.~\eqref{high-dim-ohfakd} as
\begin{equation}
\scalebox{0.91}{$
\begin{split}
 \dot{y}_i\left( \bold{x} , t\right)= \dot{X}_1^{i}(t)+ \sum_{j=1}^{d} \frac{  \left[  \bold{O}(t)~\boldsymbol{\mathcal{G}}(t)~ \bold{O}^{T}(t)  \right]_{ij}     }{2} ~\left[ y_j\left( \bold{x} , t\right)  -X_1^{j}(t) \right].
 \end{split}$}
  \label{gen-eq4}
\end{equation}
Finally, plugging this in the formula
\begin{align}
S_{\text{tot}}(t_f) &= \frac{k_B}{D} \int _0^{t_f}dt~\int _{-\infty}^{\infty}d\bold{x}~P_0(\bold{x})~\dot{\bold{y}}\left( \bold{x},t\right)^2, \label{2d-bound-high-dimension}
\end{align}
we obtain the bound in general $d$ dimensions as
\begin{equation}
\scalebox{0.85}{$
\begin{split}
S_{\text{tot}}^{12}(t_f) &= \frac{k_B}{D} \int _{0}^{t_f} dt  \left[   \dot{\bold{X}}_1(t)^2  + \frac{\text{Trace}  \Big\{  \bold{O}(t)~\boldsymbol{\mathcal{G}}^2(t)~ \bold{O}^{T}(t) ~\bold{A}(t)  \Big\}}{4} \right],  \\
 &=\frac{k_B}{D} \int _{0}^{t_f} dt  \left[   \dot{\bold{X}}_1(t)^2  + \frac{1}{4}\text{Trace}  \Big\{  \boldsymbol{\mathcal{G}}^2(t)~ \boldsymbol{\Lambda}(t)~  \Big\} \right],
\end{split}$}\label{gen-eq5}
\end{equation}
where the elements of $\boldsymbol{\mathcal{G}}(t)$ matrix are fully given in terms of $\bold{A}(t)$ in Eq.~\eqref{gen-eq3} which in turn depends on the variance and covariance of $\bold{x}(t)$. Eq.~\eqref{gen-eq5} represents a general bound on the mean entropy production in higher dimensions. For the two-dimensional case, we obtain an explicit expression (see Supplementary note 3)
\begin{equation}
\scalebox{0.91}{$
\begin{split}
& S_{\text{tot}}^{12} (t_f) = \frac{k_B}{D} \int _{0}^{t_f} dt \left[   \dot{X}_1^{1}(t) ^2 + \frac{\dot{A} _{11} (t) ^2}{4 A _{11}(t)} + \dot{X}_1^{2}(t) ^2 + \frac{\dot{A} _{22} (t) ^2}{4 A _{22}(t)}  \right.  \\
 & \left.  + \frac{\left\{ A _{12}(t) \frac{d}{dt} \left(  A _{11}(t) A _{22}(t)  \right) -2 A _{11}(t) A _{22}(t) \dot{A} _{12}(t) \right\}^2   }{ 4 A _{11}(t) A _{22}(t) \left\{ A _{11}(t)+A _{22}(t) \right\} \left\{  A _{11}(t) A _{22}(t)-A _{12}(t)^2  \right\}   }     \right]. 
 \end{split}$} 
 \label{2d-bound}
\end{equation}
The bound is equal to the sum of two one dimensional bounds in Eq.~\eqref{estimate-eq-1} corresponding to two coordinates $x_1(t)$ and $x_2(t)$ plus a part that arises due to the cross-correlation between them. In absence of this cross-correlation $ \left( A_{12}(t) = 0 \right)$, Eq.~\eqref{2d-bound} correctly reduces to the sum of two separate one dimensional bounds. \\

%\subsection{Optimal protocols with first two moments}
\noindent
\textit{Optimal protocols with first two moments: }
Let us now look at the flux and the distribution associated with this optimal dissipation. For this, we first rewrite Eq.~\eqref{high-dim-ohfakd} in vectorial form as
\begin{align}
\dot{\bold{y}} \left( \bold{x},t \right) = \dot{\bold{X}}_1(t) + \frac{\boldsymbol{\gamma}(t)}{2} \left[  \bold{y} \left( \bold{x},t \right)-  \bold{X}_1(t)  \right],\label{gen-soln-vel}
\end{align}
and solve it to obtain the solution $\bold{y} \left( \bold{x},t \right)$ as
\begin{align}
\bold{y} \left( \bold{x},t \right) = \bold{X}_1(t) + \boldsymbol{ \Omega }(t) \left[  \bold{x}-  \bold{X}_1(t)  \right]. \label{gen-soln-yd}
\end{align}
Here $\boldsymbol{ \Omega }(t)$ stands for the time-ordered exponential and is given by
\begin{align}
\boldsymbol{ \Omega }(t) & = \mathcal{T}\Big[ \exp \left(\int _{0}^{t} d t' ~ \frac{\boldsymbol{\gamma}(t')}{2} \right)\Big], \nonumber \\
& = \sum _{n=0}^{\infty} \int _{0}^{t} dt_1\int _{0}^{t_1} dt_2 \int _{0}^{t_2} dt_3...\int _{0}^{t_{n-1}} dt_n~\nonumber \\
&~~~~~~~~~~~~~~~\times \frac{\boldsymbol{\gamma}(t_1)}{2}\frac{\boldsymbol{\gamma}(t_2)}{2}....\frac{\boldsymbol{\gamma}(t_n)}{2},
\label{omega}
\end{align}
with $\boldsymbol{\gamma}(t)$ defined in Eq.~\eqref{gen-eq3} and $\mathcal{T}$ corresponding to the time-ordering operator. Finally, substituting the solution $\bold{y} \left( \bold{x},t \right)$ from Eq.~\eqref{gen-soln-yd} in Eq.~\eqref{genn-transport-map} gives the optimal flux, whereas using Eq.~(S26) in the Supplementary note, one obtains the optimal distribution
\begin{align}
\bold{v}^{12}\left( \bold{x},t \right) & = \dot{\bold{X}}_1(t) + \frac{\boldsymbol{\gamma}(t)}{2} \left[   \bold{x}-  \bold{X}_1(t)  \right], \\
P^{12}\left( \bold{x},t \right) & = \frac{P_0 \Big( \bold{X}_1(t) +  \boldsymbol{ \Omega }^{-1}(t) \left[   \bold{x}-  \bold{X}_1(t)  \right] \Big)}{|\text{det} \left( \boldsymbol{ \Omega }(t) \right)|}.
\end{align}
Now the optimal force-field follows straightforwardly from Eq.~\eqref{force-eqqvd} as
\begin{align}
\bold{F}\left( \bold{x},t \right) &= \frac{k_B T}{D} \left[  \dot{\bold{X}}_1(t) + \frac{\boldsymbol{\gamma}(t)}{2} \Bigg( \bold{x}-  \bold{X}_1(t)  \Bigg) \right. \nonumber \\
& ~~~~~\left. +D~ \boldsymbol{\nabla} \ln   P_0 \Big( \bold{X}_1(t) +  \boldsymbol{ \Omega }^{-1}(t) \left[   \bold{x}-  \bold{X}_1(t)  \right] \Big)    \right].
\end{align}
To sum up, we have derived a lower bound to the total dissipation $S_{\text{tot}}(t_f)$ in general $d$ dimensions just from the information about first two moments. Together with this, we also calculated the force-field that gives rise to this optimal value. In what follows, we consider the most general case where we know the first-$m$ (mixed) moments of the state variable $\bold{x}(t) = \{x_1(t),x_2(t),...,x_d(t) \}$. \\

%\subsection{Fixing first $m$ moments in general $d$ dimensions}
\noindent
\textit{Fixing first $m$ moments in general $d$ dimensions: }
Let us define a general mixed moment as
\begin{align}
X_{m}^{i_1, i_2,..,i_m}(t) = \int _{-\infty}^{\infty} d \bold{x}~P \left( \bold{x},t \right)~ x_{i_1} x_{i_2}....x_{i_3}, \label{gen-m-point-d-eq-1}
\end{align}
%\begin{align}
%\mathcal{C}_{l_1, l_2,..,l_d}(t) = \int _{-\infty}^{\infty} d \bold{x}~P \left( \bold{x},t \right)~ \left(x_1 \right)^{l_1} \left(x_2 \right)^{l_2}...\left(x_d \right)^{l_d}, \label{gen-m-point-d-eq-1}
%\end{align}
where $i _{j} \in \mathbb{Z}^+$ and $i_1 \geq i_2 \geq i_3 \geq..\geq i_m$ for all $1 \leq i_{l} \leq d$. For $m=1$ and $m=2$, Eq.~\eqref{gen-m-point-d-eq-1} reduces to first two moments in Eq.~\eqref{bxgapw}. Here, we are interested in the general case for which we consider the following action
\begin{equation}
\scalebox{0.93}{$
\begin{split}
& \mathbb{S}\left( \bold{x}, \bold{v}, t_f\right) = \frac{D}{k_B} S _{\text{tot}}(t_f) + \int _{0}^{t_f}dt \left[\sum _{l=1}^{m} \sum _{~\{i_j \}=1}^{d}\mu _{i_1, i_2,...,i_l}(t) \right.~
 \\
& \Bigg.\times X_{l}^{i_1, i_2,..,i_l}(t) \Bigg]+ \sum _{l=1}^{m} \sum _{~\{i_j \}=1}^{d}~\zeta _{i_1, i_2,...,i_l}(t_f)~X_{l}^{i_1, i_2,..,i_l}(t_f). 
\end{split}$}
\label{gen-m-point-d-eq-2}
\end{equation}
%\small{\begin{align}
%\mathbb{S}\left( \bold{x}, \bold{v}, t_f\right) &= \frac{D}{k_B} S _{\text{tot}}(t_f) + \int _{0}^{t_f}dt \Bigg[  \sum _{l_1, l_2,..,l_d} \Theta \Big( m-l_1-l_2-..-l_d\Big) ~\mu _{l_1,l_2,...,l_d}(t)~\mathcal{C}_{l_1, l_2,..,l_d}(t)   \Bigg] \nonumber \\
%& ~~~~~~~~ + \sum _{l_1, l_2,..,l_d} \Theta \Big( m-l_1-l_2-..-l_d\Big) ~\zeta _{l_1,l_2,...,l_d}(t_f)~\mathcal{C}_{l_1, l_2,..,l_d}(t_f). \label{gen-m-point-d-eq-2}
%\end{align}}
%The notation $\Theta(z)$ stands for the Heaviside step function with $\Theta(z) = 1$ if $z \geq 0$ and $\Theta(z) = 0$ otherwise. The presence of this function ensures that we only fix at most $m$-moments. 
In conjunction to the previous sections, we again recast this action in terms of $\bold{y} \left( \bold{x},t \right)$ defined in Eq.~\eqref{genn-transport-map} as
\begin{align}
\begin{split}
\mathbb{S}&\left( \bold{y}, \dot{\bold{y}}, t_f\right) \\
& =\int _0 ^{t_f} dt \int _{-\infty}^{\infty} d \bold{x}~ P_0 \left( \bold{x} \right) \left[ \dot{\bold{y}}\left( \bold{x},t \right) ^2 \right. \\
& \quad + \sum _{l=1}^{m} \sum _{~\{i_j \}=1}^{d}~\mu _{i_1, i_2,...,i_l}(t)  \left\{ \prod _{q=1}^{l} y _{i_q}\left( \bold{x},t \right) \right\} \bigg] 
\end{split} \nonumber \\
   &+ \int _{-\infty}^{\infty} d \bold{x}~ P_0 \left( \bold{x} \right)\sum _{l=1}^{m} \sum _{~\{i_j \}=1}^{d}~\zeta _{i_1, i_2,...,i_l}(t_f) \prod _{q=1}^{l} y _{i_q}\left( \bold{x},t_f \right) . \label{gen-m-point-d-eq-3}
\end{align}
%\begin{align}
%\mathbb{S}\left( \bold{y}, \dot{\bold{y}}, t_f\right)  & =\int _0 ^{t_f} dt \int _{-\infty}^{\infty} d \bold{x}~ P_0 \left( \bold{x} \right) \left[ \dot{\bold{y}}\left( \bold{x},t \right) ^2   + \sum _{\boldsymbol{l}} \text{}^{(m)}  ~\mu _{l_1,l_2,...,l_d}(t)~  \left\{\prod _{i=1}^{d}y_i\left( \bold{x}, t \right)^{l_i} \right\}
%\right] \nonumber \\
%& ~~~~~~~+\int _{-\infty}^{\infty} d \bold{x}~ P_0 \left( \bold{x} \right)~\sum _{\boldsymbol{l}} \text{}^{(m)}  ~\zeta _{l_1,l_2,...,l_d}(t_f)~ \left\{\prod _{i=1}^{d}y_i\left( \bold{x}, t_f \right)^{l_i} \right\}, \label{gen-m-point-d-eq-3}
%\end{align}
%where we have used the following short-hand notation for the summation
%\begin{align}
%\sum _{l_1, l_2,..,l_d} \Theta \Big( m-l_1-l_2-..-l_d\Big) \cdot= \sum _{\boldsymbol{l}} \text{}^{(m)}\cdot
%\end{align}
Optimising this action in the same way as before, we obtain the Euler-Lagrange equation
%\begin{align}
%2 \ddot{y}_i\left( \bold{x}, t \right)& =\sum _{\boldsymbol{l}} \text{}^{(m)}\Bigg[ \left\{ \frac{ \mu _{l_1,l_2,...,l_d}(t)}{l_i^{-1}~y_i\left( \bold{x}, t \right)} \right\}  \prod_{j=1}^{d}  y_j\left( \bold{x}, t \right)^{l_j} \Bigg], \label{gen-m-point-d-eq-4}
%\end{align}
\begin{equation}
\scalebox{0.91}{$
\begin{split}
2 \ddot{y}_i\left( \bold{x}, t \right)& =\sum _{l=1}^{m} \sum _{~\{i_j \}=1}^{d}~\mu _{i, i_2,...,i_l}(t) ~y _{i_2}\left( \bold{x},t \right) y _{i_3}\left( \bold{x},t \right)...y _{i_l}\left( \bold{x},t \right) ,
\end{split}$}
\label{gen-m-point-d-eq-4}
\end{equation}
along with the boundary conditions 
\begin{align}
2 \dot{y}_i\left( \bold{x}, t_f \right)& =-\sum _{l=1}^{m} \sum _{~\{i_j \}=1}^{d}~y _{i_2}\left( \bold{x},t_f \right) y _{i_3}\left( \bold{x},t_f \right)...y _{i_l}\left( \bold{x},t_f \right) \nonumber \\
& ~~~~~~~~~~~~~~~~~~~~~~~~~~\times \zeta _{i, i_2,...,i_l}(t_f) , \label{gen-m-point-d-eq-5} \\
y_i\left( \bold{x}, 0 \right) & = x_i. \label{gen-m-point-d-eq-6}
\end{align}
The Lagrange multipliers $\mu _{i_1,i_2,...,i_l}(t)$ and $\zeta _{i_1,i_2,...,i_l}(t)$ can be computed using time-derivatives of the (mixed) moments $X_{l}^{i_1, i_2,..,i_l}(t) $ as
\footnotesize{\begin{align}
\dot{X}_{r}^{i_1, i_2,..,i_r}(t)& = -\frac{1}{2} \sum _{k=1}^{r} \sum _{l=1}^{m} \sum _{~\{i_j' \}=1}^{d}\Bigg[\zeta _{i_k, i_2',i_3',...,i_l'}(t) \Bigg. \nonumber \\
&\Bigg.\times  X_{r+l-2}^{i_1, i_2,...,i_{k-1},i_{k+1}.,i_r, i_2'.,i_3',...,i_l'}(t) \Bigg],
\label{gen-m-point-d-eq-7} \\
\ddot{X}_{r}^{i_1, i_2,..,i_r}(t) &= \frac{1}{2} \sum _{k=1}^{r} \sum _{l=1}^{m} \sum _{~\{i_j' \}=1}^{d}\Bigg[ \mu _{i_k, i_2',i_3',...,i_l'}(t) \Bigg.~ \label{gen-m-point-d-eq-8} \\
 \Bigg.\times  X_{r+l-2}&^{i_1, i_2,...,i_{k-1},i_{k+1}.,i_r, i_2'.,i_3',...,i_l'}(t) \Bigg] +\mathcal{B}_{i_1, i_2,..,i_r}(t) , \nonumber \\
\text{with }\mathcal{B}_{i_1, i_2,..,i_r}(t)  & = \int _{-\infty}^{\infty} d \bold{x}~ P_0 \left( \bold{x} \right)  \left[ \prod _{j=1}^{r}y_{i_j} \left( \bold{x}, t \right) \right] \nonumber \\
& ~~~~~~~~~~~~~~~\times \sum _{k ,l=1 \atop k \neq l}^{r}~\frac{\dot{y}_{i_l} \left( \bold{x}, t \right)~\dot{y}_{i_k} \left( \bold{x}, t \right)}{y_{i_l} \left( \bold{x}, t \right) ~y_{i_k} \left( \bold{x}, t \right)}. \label{gen-m-point-d-eq-9} 
\end{align}}
%\begin{align}
%\dot{X}_{r}^{i_1, i_2,..,i_r}(t) &= -\frac{1}{2} \sum _{k=1}^{r} \sum _{l=1}^{m} \sum _{~\{i_j' \}=1}^{d}~\zeta _{i_k, i_2',i_3',...,i_l'}(t)~ X_{r+l-2}^{i_1, i_2,...,i_{k-1},i_{k+1}.,i_r, i_2'.,i_3',...,i_l'}(t)
%\label{gen-m-point-d-eq-7} \\
%\ddot{X}_{r}^{i_1, i_2,..,i_r}(t) &= \frac{1}{2} \sum _{k=1}^{r} \sum _{l=1}^{m} \sum _{~\{i_j' \}=1}^{d}~\mu _{i_k, i_2',i_3',...,i_l'}(t)~ X_{r+l-2}^{i_1, i_2,...,i_{k-1},i_{k+1}.,i_r, i_2'.,i_3',...,i_l'}(t) \nonumber \\
%& ~~~~~~~~~~~~~~  +\mathcal{B}_{i_1, i_2,..,i_r}(t) , \label{gen-m-point-d-eq-8} \\
%\text{with }\mathcal{B}_{i_1, i_2,..,i_r}(t)  & = \int _{-\infty}^{\infty} d \bold{x}~ P_0 \left( \bold{x} \right) \sum _{k ,l=1 \atop k \neq l}^{r}~\frac{\dot{y}_{i_l} \left( \bold{x}, t \right)~\dot{y}_{i_k} \left( \bold{x}, t \right)}{y_{i_l} \left( \bold{x}, t \right) ~y_{i_k} \left( \bold{x}, t \right)}  \prod _{j=1}^{r}y_{i_j} \left( \bold{x}, t \right). \label{gen-m-point-d-eq-9} 
%\end{align}
\normalsize{Once} again, for $m>2$, we find dependence of the right hand side of Eqs.~\eqref{gen-m-point-d-eq-7} and \eqref{gen-m-point-d-eq-8} on (mixed) moments with order higher than $m$. As discussed before, one can tackle this by  utilizing the solution $y_i\left( \bold{x},t \right)$ recursively to obtain these higher moments [as discussed after Eq.~\eqref{gen-lin-eqdfa-2}].

Solving Eq.~\eqref{gen-m-point-d-eq-4} numerically yields the solution $\bold{y}_{*}\left( \bold{x},t \right)$ using which in Eq.~\eqref{2d-bound-high-dimension} we obtain a lower bound to $S_{\text{tot}}(t_f)$. The associated optimal flux and optimal distribution can also be calculated as before.

\section*{Results}
%\section*{Theoretical examples}
%\label{application-sec}
\noindent
\textbf{Theoretical examples.} We now test our framework on two analytically solvable toy models. The first one consists of a one dimensional free diffusion model with initial position drawn from the distribution $P_0(x_0) \sim \exp(-x_0^4)$. This example will demonstrate how our method becomes more accurate when we incorporate knowledge of higher moments. The second example involves two-dimensional diffusion in a moving trap, serving as an illustration of our method's applicability in higher dimensions.\\

%\subsection{Free diffusion with $P_0(x_0) =  \exp(-x_0^4) / 2 \Gamma(5/4)$}
%\label{sec-free-diffusion}
\noindent
\textit{Free diffusion:} We first consider a freely diffusing particle in one dimension 
\begin{align}
\frac{dx}{dt} = \sqrt{2 D}~ \eta(t),~~\text{with }P_0(x_0) =  \frac{e^{-x_0^4}}{2 \Gamma(5/4)}. \label{model-1}
\end{align}
This example serves two purposes: First, we can obtain the exact expression of $S_{\text{tot}}(t_f)$ which enables us to rigorously compare the bounds derived above with their exact counterpart. Second, we discuss the intricacies that arise when we fix more than first two moments. To begin with, the position distribution $P(x,t_f)$  is given by
\begin{align}
P(x,t_f) = \frac{1}{\sqrt{4 \pi D t_f}}~\int _{-\infty}^{\infty} dx_0~ P_0(x_0)~ e^{   -\frac{(x-x_0)^2}{4 Dt_f} }. \label{dist-model-1}
\end{align}
Since no external force acts on the system, the total entropy produced will be equal to the total entropy change of the system. Therefore, the mean $S_{\text{tot}}(t_f)$ reads
\begin{align}
S_{\text{tot}}(t_f) =&  -k_B \int _{-\infty} ^{\infty} dx~ P(x,t_f)~ \log P(x,t_f) \nonumber \\
& ~~~~~~+ k_B \int _{-\infty} ^{\infty} dx_0~ P_0(x_0) ~\log P_0(x_0). \label{tentropy-model1}
\end{align}
We emphasize that Eqs.~\eqref{dist-model-1} and \eqref{tentropy-model1} are exact results and do not involve  any approximation. 

\begin{figure*}[t]
	\includegraphics[scale=0.4]{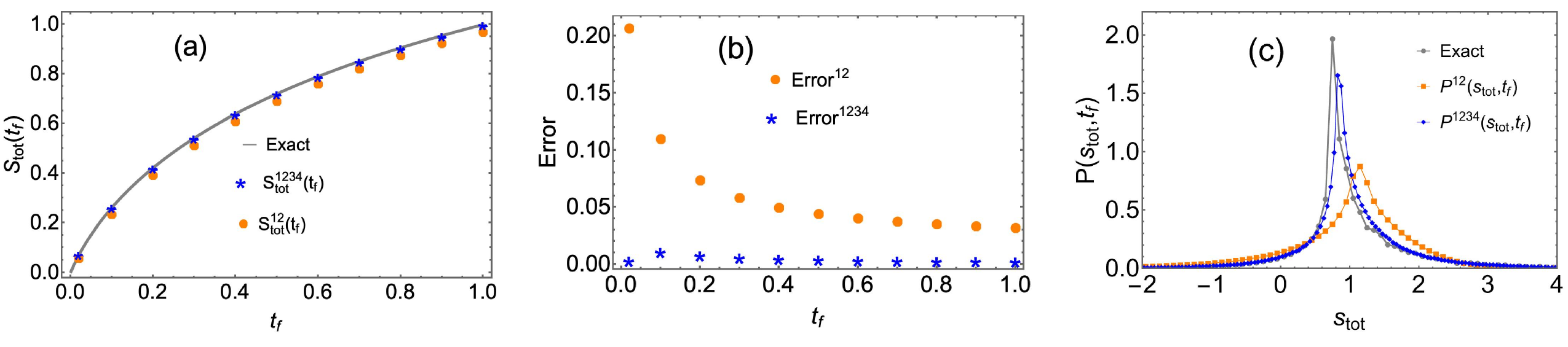}
	\centering
	\caption{\textbf{Estimation for the free diffusion model.} (a) Estimation of the mean total entropy produced $S_{\text{tot}}(t_f)$ for model \eqref{model-1} by using the first two moments (in orange) and first four moments (in blue). The gray curve is the exact result whose expression is given in Eq.~\eqref{tentropy-model1}. \bluew{(b) The relative error in the estimated values compared to the exact value as a function of $t_f$ [see Eq.~\eqref{error-new-1}]. Orange symbols represent error due to the first two moments while blue ones correspond to the first four moments.} (c) Probability distribution of the total entropy produced $s_{\text{tot}}(t_f)$ for the optimal protocol obtained by fixing the first two moments (in orange) and first four moments (in blue) till time $t_f =1$. Here also, the gray symbols correspond to the simulation of Eq.~\eqref{model-1}. For both panels, we have used $D=1\text{ and }k_B=1$}
	\label{fig-entropy-model-1}
\end{figure*}

Let us now analyse how our bound compares with this exact value. The first four moments of the position are given by
\begin{align}
& X_1(t) = X_3(t)= 0, \label{trshjgd-1} \\
& X_2(t) = 2 D t+\frac{\Gamma (3/4)}{ \Gamma (1/4)}, \label{trshjgd-111}\\
& X_4(t) = \frac{1}{4} + 12 ~\frac{\Gamma (3/4)}{ \Gamma (1/4)} Dt + 12 D^2 t^2. \label{trshjgd-2}
\end{align}
Combining this with Eq.~\eqref{estimate-eq-2}, shows that the dissipation bound for the first two moments reads
\begin{align}
S_{\text{tot}}^{\text{12}}(t_f)& = \int _0^{t_f} dt \frac{k_B D}{2 Dt + \frac{\Gamma (3/4)}{ \Gamma (1/4)}} , \nonumber \\
& = \frac{k_B}{2} \log \left( \frac{2 D t_f + \frac{\Gamma (3/4)}{ \Gamma (1/4)}}{\frac{\Gamma (3/4)}{ \Gamma (1/4)}}\right). \label{avhqob1}
\end{align}
As illustrated in Figure \ref{fig-entropy-model-1}(a), this bound $S_{\text{tot}}^{\text{12}}(t_f)$, although not exact, is still quite close to the exact value in Eq.~\eqref{tentropy-model1}. This is somewhat surprising as one can show that the force field that saturates the bound is given by
\begin{equation}
    F^{12}(x,t) = k_B T \left[  \frac{x}{2 D t +    \frac{\Gamma \left( 3/4\right)}{\Gamma \left( 1/4\right)} }   - \frac{4 x^3}{\alpha(t)^4} \right], \label{diff-f12} 
\end{equation}
which is significantly different from the force field associated with free relaxation ($F(x,t)=0$). The error is shown in Figure \ref{fig-entropy-model-1}(b) (discussed later).

Going beyond mean, we have also plotted the full probability distribution of total dissipation in Figure \ref{fig-entropy-model-1}(c). To do this, we evolve the particle with the optimal force-field in Eq.~\eqref{diff-f12} starting from the position $x_0$ drawn from the distribution $P_0(x_0)$ in Eq.~\eqref{model-1}. For different trajectories thus obtained, we measure the total entropy produced and construct the probability distribution out of them. Figure \ref{fig-entropy-model-1}(c) shows the result. Here, we see deviation from the exact result indicating that fixing first two moments only does not give a very good approximation of dissipation beyond its average value. To improve this, we instead fix the first four moments of position and carry out this analysis.

%with associated optimal protocols and time-dependent distribution given by
%\begin{align}
%P^{12}(x,t) & = \frac{e^{  - \frac{x^4}{\alpha (t)^4}}}{2 \Gamma \left(  \frac{5}{4} \right)~\alpha(t)},~\alpha(t) = \sqrt{\frac{\frac{\Gamma \left( \frac{3}{4}\right)}{\Gamma \left( \frac{1}{4}\right)}}{2 Dt + \frac{\Gamma \left( \frac{3}{4}\right)}{\Gamma \left( \frac{1}{4}\right)}}} \label{diff-p12}\\
%v^{12}(x,t) & = \frac{Dx}{2D t + \frac{\Gamma \left( 3/4\right)}{\Gamma \left( 1/4\right)}}. \label{diff-v12} \\
%\end{align}
%\redw{Notice that although the actual model consists of a freely diffusing particle, the optimal force-field is significantly different and possesses rather a non-trivial form. This is crucial for obtaining the optimal total dissipation.}

One can wonder whether it is possible to get an even better estimate by taking into account the first four moments.
To check this, we will now proceed by solving the Euler-Lagrange equation \eqref{euler-gen-eq-1} numerically for $m=4$. Since, this is a second-order differential equation, we need two boundary conditions in time. One of them is Eq.~\eqref{euler-gen-eq-3} which gives $y(x,0) = x$. However, the other one in Eq.~\eqref{euler-gen-eq-2} gives a condition at final time $t_f$ which is difficult to implement numerically. For this, we use the following trick.
We expand the initial $\dot{y}(x,0)$ as
\begin{align}
\dot{y}(x,0) = \theta _0  + \theta _1 x+....+\theta _s x^s, 
%\dot{y}(x,0) = \theta _0 T_0(x) + \theta _1 T_1(x)+....+\theta _s %T_{s}(x), 
\label{ini-ydot-eqq}
\end{align}
where we typically choose $s = 9$. The coefficients $\theta _i$ are free to vary but must give the correct time derivative of first four moments in Eqs.~\eqref{trshjgd-1}-\eqref{trshjgd-2}. We then assign a cost function as 
\begin{equation}
\scalebox{0.91}{$
\begin{split}
\mathcal{C}(t_f) = \int _{-\infty}^{\infty} dx~P_0(x)~\left[  2 \dot{y}(x, t_f)  +\sum _{i=1}^{4}i~ \zeta _i(t_f) ~y(x,t_f)^{i-1} \right]^2.
\end{split}$} 
\end{equation}
Observe that the cost function vanishes if the boundary condition \eqref{euler-gen-eq-2} is satisfied. Starting from $y(x,0) = x$ and $\dot{y}(x,0)$ in Eq.~\eqref{ini-ydot-eqq}, we solve the Euler-Lagrange equation \eqref{euler-gen-eq-1} numerically and measure the cost function at the final time. We then perform a multi-dimensional gradient descent in $\theta$-parameters to minimise $\mathcal{C}(t_f)$. Eventually, we obtain the numerical form of the optimal map $y_{*}(x,t)$.

Using this map, we then obtain a refined bound on the dissipation, $S_{\text{tot}}^{1234}(t_f)$. This is illustrated in Figure \ref{fig-entropy-model-1}(a). As seen, this bound is better than $S_{\text{tot}}^{12}(t_f)$ and in fact, matches very well with the exact one. \bluew{To exemplify the tightness of these bounds, we have plotted, in Figure \ref{fig-entropy-model-1}(b), the relative error
\begin{align}
\text{error} = \frac{\text{exact value}-\text{estimated value}}{\text{exact value}}, \label{error-new-1}
\end{align}
as a function of $t_f$. This clearly shows that the estimate with four moments is tighter than the one with two moments. We also observe that the tightness of bound increases as we increase the final time $t_f$. With larger $t_f$, the final distribution $P(x,t_f)$ in Eq.~\eqref{dist-model-1} approaches a Gaussian form. For instance $X_2(t) \simeq 2 D t$ and $X_4(t) \simeq 12 D^2 t^2$ at large times, exhibiting Gaussian characteristics. Therefore, for large $t_f$, our method becomes exact.} Furthermore, we can also obtain the associated optimal protocols by numerically inverting Eqs.~\eqref{Lag-cord-eq-1} and \eqref{OTT-dist}. With these protocols, we then obtain the distribution of total entropy produced which is shown in Figure \ref{fig-entropy-model-1}(c). Compared to the previous case, we again find that the estimated distribution is now closer to the exact one.

\bluew{In this example, we were able to calculate the moments analytically and obtain the exact lower bound either analytically (for two moments) or numerically (otherwise). However, in many cases, it is not possible to calculate the moments  and one needs to rely on the available trajectories to obtain them. An example of this is presented below. Now if the number of trajectories (denoted by $R$) is large, then one still has an exact numerical estimate of the moments and hence the exact lower bound. However, with small $R$, the moments deviate from their actual values and this affects the accuracy of our estimates. Moreover, the effect of noise will be greater on higher moments compared to the lower moments. This means that estimate with higher moments is more likely to be inaccurate than with lower moments for small trajectory number. On the other hand, when $R$ is significantly large, the accuracy of estimate with higher moments is always high. In Supplementary note 4, we have demonstrated the effect of noise arising due to the finite number of trajectories on the lower bound.} Similarly, while acquiring the trajectory data experimentally, there are intrinsic errors that occur due to the data measurement. Due to this, the observed position is different from the actual position and one incurs some error during the measurement process \citep{metzler}. In Supplementary note 5, we have discussed the impact of such measurement errors on our lower bound. \\

%\subsection{Two dimensional Brownian motion}

\noindent
\textit{Two dimensional Brownian motion:} The previous example illustrated our method in a one-dimensional setting. In this section, we look at a two-dimensional system with position $\bold{x}(t) = \left( x_1(t), x_2(t) \right)$. The system undergoes motion in presence of a moving two-dimensional harmonic oscillator $U(x_1, x_2,t) = K \left (x_1-v t \right)\left (x_2-v t \right)$ and its dynamics is governed by
\begin{align}
&\frac{dx_1}{dt} = -K(x_2-v t)+\sqrt{2 D} ~\eta_1(t), \\
&\frac{dx_2}{dt} = -K(x_1-v t)+\sqrt{2 D} ~\eta_2(t).
\end{align}
We also assume that the initial position is drawn from Gaussian distribution as
\begin{align}
P_0(x_1(0),x_2(0)) = \frac{1}{2 \pi b_0^2} ~\exp \left(-\frac{x_1(0)^2+x_2(0)^2}{2 b_0^2}\right).
\end{align}
For this example, our aim is to compare the bound derived in Eq.~\eqref{2d-bound} with the exact expression. Using the Langevin equation, we obtain the first two cumulants and covariance to be
\begin{align}
& X_1^{1}(t) = X_1^{2}(t) = X_1(t)=\frac{v}{K} \left(Kt-1+e^{-K t} \right) , \\
&A_{11}(t) = A_{22}(t) = b_0^2 \cosh (2 K t)+\frac{D}{K} \sinh (2 K t), \\
& A_{12}(t) = A_{21}(t) = -b_0^2 \sinh (2 K t)+\frac{D}{K} \left[ 1-\cosh (2 K t)  \right].
\end{align} 
Plugging them in Eq.~\eqref{2d-bound} gives
\footnotesize{\begin{align}
&\frac{D S^{12}_{\text{tot}}}{k_B}\left( t_f  \right)  \nonumber  \\
& =   \int _{0}^{t_f} dt \left[  2 \dot{X}_1(t)^2 + \frac{\Big( \dot{A}_{11}(t)+\dot{A}_{12}(t)  \Big)^2 }{4 \Big(  A_{11}(t)+A_{12}(t)    \Big)} + \frac{\Big( \dot{A}_{11}(t)-\dot{A}_{12}(t)  \Big)^2 }{4 \Big(  A_{11}(t)-A_{12}(t)    \Big)}   \right] ,  \\
& =  \int _{0}^{t_f} dt\left[  2 v^2(1-e^{-Kt})^2 + \frac{ \left(K b_0^2 - D \right)^2 e^{-4 K t}}{b_0^2 e^{-2 K t}+\frac{D}{K} \left(1-e^{-2K t} \right)} \right. \nonumber \\
&~~~~~~~~~~~~~~~~~~~~~~~~~~ \left.+\frac{ \left(K b_0^2 + D \right)^2 e^{4 K t}}{b_0^2 e^{2 K t}+\frac{D}{K} \left(e^{2K t}-1 \right)}  \right]. 
\label{HD-ent-ex}
\end{align}}
\normalsize{In} fact, one can carry out exact calculations for this model. To see this, let us introduce the change of variable
\begin{align}
R_1 = \frac{x_1 + x_2}{\sqrt{2}},~~~~~~\text{and}~~~~~R_2 = \frac{x_1 -x_2}{\sqrt{2}}
\end{align}
and rewrite the Langevin equations as
\begin{align}
&\frac{dR_1}{dt} = -K\left(R_1-\sqrt{2} v t \right)+\sqrt{2 D} ~\eta_{R_1}(t), \\
&\frac{dR_2}{dt} =K R_2+\sqrt{2 D} ~\eta_{R_2}(t),
\end{align}
\begin{figure*}[t]
  \centering
  \includegraphics[scale=0.5]{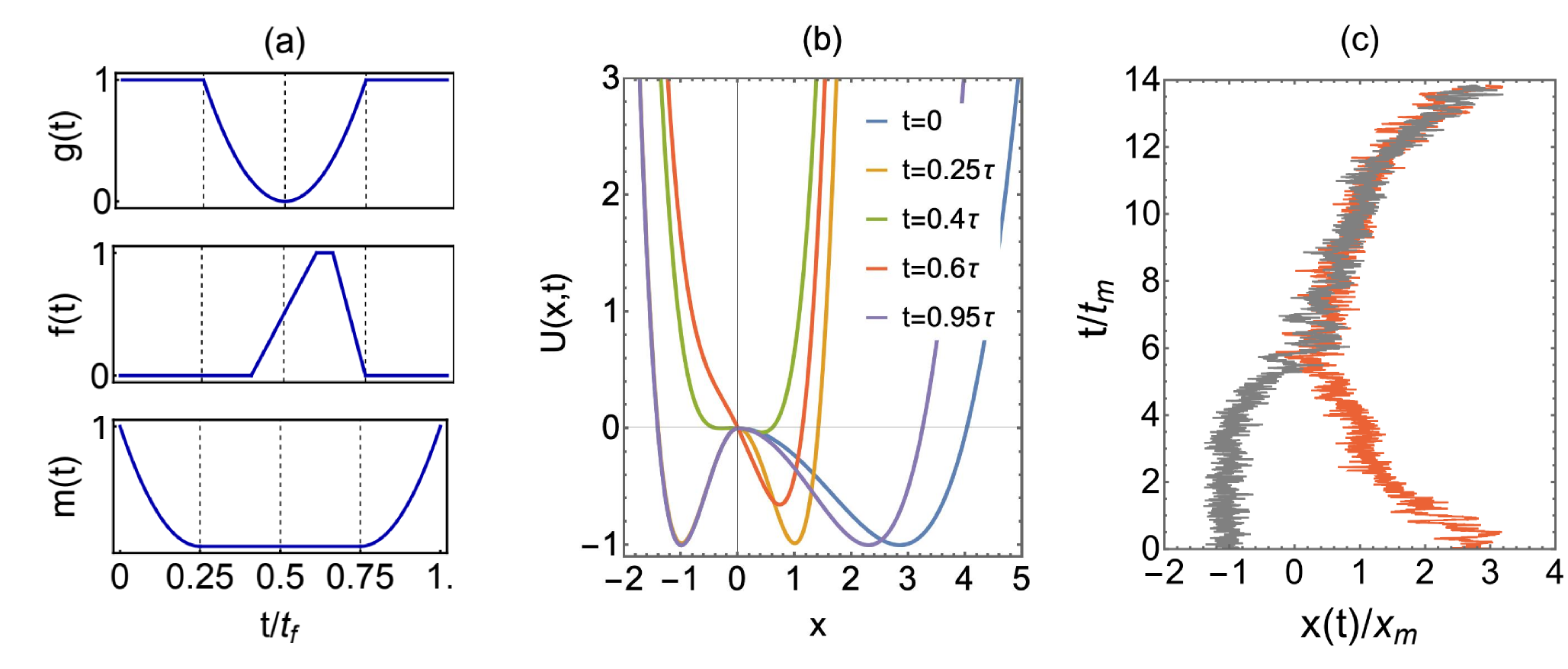}
\centering
\caption{\textbf{Details of the bit erasure experiment.} (a) Plot of the the control functions, $g(t)$, $f(t)$ and $m(t)$, that are modulated experimentally to carry out the erasure operation. (b) Time modulation of the potential $U(x,t)$ in Eq.~\eqref{BE-eq-2} at different times. (c) Experimental trajectories obtained through by changing the protocol. The grey trajectory starts from the left well initially and ends up in the right well at the end of the cycle. On the other hand, the red trajectory starts from the right well and also ends up in the right well.}    
\label{BE-fig-1} 
\end{figure*}
where $\eta_{R_1}(t)$ and $\eta_{R_2}(t)$ are two independent Gaussian white noises with zero mean and delta correlation. Clearly these two equations are independent of each other which means that we can treat $R_1(t)$ and $R_2(t)$ as two independent one dimensional Brownian motions. Also, observe that both of them are Gaussian processes. Following our analysis in Supplementary note 2, we can now express average dissipation completely in terms of first two cumulants. The first two cumulants for our model are
\begin{align}
& \langle R_1(t) \rangle = \frac{\sqrt{2}v}{K} \left( Kt -1 +e^{-K t} \right),~~~~~\langle R_2(t) \rangle = 0,  \\
& A _{R_1}(t) = \langle R_1^2(t) \rangle-\langle R_1(t) \rangle^2 = b_0^2 ~e^{-2 K t} + \frac{D}{K} \left( 1-e^{-2 K t}  \right), \\
& A _{R_2}(t) = \langle R_2^2(t) \rangle-\langle R_2(t) \rangle^2 = b_0^2 ~e^{2 K t} + \frac{D}{K} \left( e^{2 K t}-1  \right).
\end{align}
Finally inserting them in Eq.~\eqref{estimate-eq-1}, we obtain
\begin{equation}
\scalebox{0.91}{$
\begin{split}
S_{\text{tot}}\left( t_f \right) & = \frac{k_B}{D} \int _ 0^{t_f} dt \left[   \left( \frac{d \langle R_1(t) \rangle }{dt}\right)^2 + \frac{\dot{A} _{R_1}(t) ^2 }{4 A _{R_1}(t) }   + \frac{\dot{A} _{R_2}(t) ^2 }{4 A _{R_2}(t) }   \right],  \\
&=\frac{k_B}{D} \int _{0}^{t_f} \left[  2 v^2(1-e^{-Kt})^2 + \frac{ \left(K b_0^2 - D \right)^2 e^{-4 K t}}{b_0^2 e^{-2 K t}+\frac{D}{K} \left(1-e^{-2K t} \right)} \right. \\ 
&~~~~~~~~~~~~~~~~~~~~~\left.+\frac{ \left(K b_0^2 + D \right)^2 e^{4 K t}}{b_0^2 e^{2 K t}+\frac{D}{K} \left(e^{2K t}-1 \right)}  \right],
\end{split}$} 
\end{equation}
which matches with the estimated entropy production in Eq.~\eqref{HD-ent-ex} exactly. In other words, our method becomes exact in this case.\\

%\section*{Bit erasure}
%\label{sec-BE}
\noindent
\textbf{Bit erasure.} Motivated by the very good agreement between the bound and the actual entropy production in simple theoretical models, we will now turn to experimental data of a more complicated system, namely bit erasure \citep{Landauer1961}.
The state of a bit can often be described by a one-dimensional Fokker-Planck equation, with a double-well potential energy landscape. This landscape can for example be produced by an optical tweezer \citep{Berut2012}, a feed-back trap \citep{Bit-1}, or a magnetic field \citep{Hong2016, Martini2016}. Initially, the colloidal particle can either be in the left well (state $0$) or in the right well (state $1$) of the double-well potential. By modulating this potential, one can assure that the particle always ends up the right well (state $1$). During this process, the expected amount of work put into the system is always bounded by the Landauer’s limit, $\langle W \rangle \geq k_B T \ln 2$ \citep{Bit-1}. Over the last few years, Landauer’s principle has also been tested in a more complex situation where symmetry between two states is broken \citep{Bit-2, Bit-3}. To add an extra layer of complexity, we consider here the scenario where this symmetry is broken by deploying an asymmetric double-well potential and we will use the data from \citep{Bit-2} to test our bound. The system satisfies the Fokker-Planck equation
\begin{align}
\frac{\partial  P(x,t)}{\partial t}  = \frac{D}{k_B T} \frac{\partial}{\partial x}\left[ P(x,t) ~\frac{\partial U(x,t)}{\partial x}  +k_B T ~ \frac{\partial  P(x,t) }{\partial x} \right],
\label{BE-eq-1}
\end{align}
where $U(x,t)$ is the time-dependent asymmetric double-well potential given by
\begin{align}
U(x,t) &= 4 E_b \left[ -\frac{ g(t)}{2} \tilde{x}^2 +\frac{\tilde{x}^4}{4}-A f(t)~\tilde{x} \right], \label{BE-eq-2} \\
\text{with } \tilde{x}= & \begin{cases} \frac{x}{x_m}~~~~~~~~~~~~~~~~~~~~~~~~~~~~~~\text{if }x<0,\\
\frac{x}{x_m}\left[1 +m(t) (\eta -1) \right]^{-1}~~~~\text{if }x \geq 0.
\end{cases}
\label{BE-eq-3}
\end{align}
Notice that initially, the potential $U(x,t)$ has its minimum located at $-x_m$ and $+\eta x_m$, where $\eta \geq 1$ is the asymmetry factor. For $\eta =1$, the potential is completely symmetric initially. On the other hand, for $\eta \neq 0$, the two wells are asymmetric and the system is out of equilibrium initially. Furthermore, the functions $m(t)$, $g(t)$ and $f(t)$ represent the experimental protocols to modulate $U(x,t)$ during erasure operation. They are defined in Figure \ref{BE-fig-1}(a) as well as in Supplementary note 6.
The experimental protocol consists of symmetrizing $U(x,t)$ by changing the function $m(t)$, while the other two functions $g(t)$ and $f(t)$ are kept fixed. Then, $g(t)$ and $f(t)$ are changed during which the barrier is lowered, the potential is tilted and again brought back to its symmetric form at time $t = 3 t_f /4 $. By this time, the particle is always at the right well. Finally, the right well is expanded to its original size with minimum located at $+ \eta x_m$ at the end of the cycle time $t_f$. The mathematical forms of these protocol functions are also given in supplementary table~(I). The time-modulation of the potential is shown in the middle panel in Figure \ref{BE-fig-1}(b) and the experimental trajectories thus obtained are shown in Figure \ref{BE-fig-1}(c).

%\begin{figure}[t]
%	\includegraphics[scale=0.3]{BE-mom1.pdf}
%	\includegraphics[scale=0.31]{BE-mom2.pdf}
%	\centering
%	\caption{First two moments associated with the erasure operation with the cycle time $t_f/t_m = 14.89$. The measured moments are fitted by the Chebyshev polynomials as indicated in Eqs.~\eqref{BE-eq-4} and \eqref{BE-eq-5}.}
%	\label{BE-fig-2}
%\end{figure} 

\begin{figure}[]
    \centering
	\includegraphics[scale=1.5]{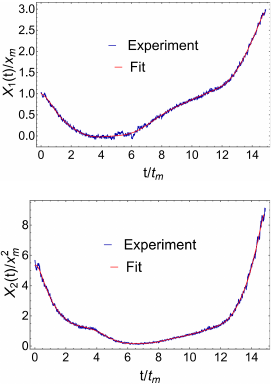}
	\centering
	\caption{\textbf{Moments from experimental trajectories.} First two moments associated with the erasure operation with the cycle time $t_f/t_m = 14.89$. The measured moments are fitted by the Chebyshev polynomials as indicated in Eqs.~\eqref{BE-eq-4} and \eqref{BE-eq-5}.}
	\label{BE-fig-2}
\end{figure} 
\begin{figure}[]
	\centering
   \includegraphics[scale=0.33]{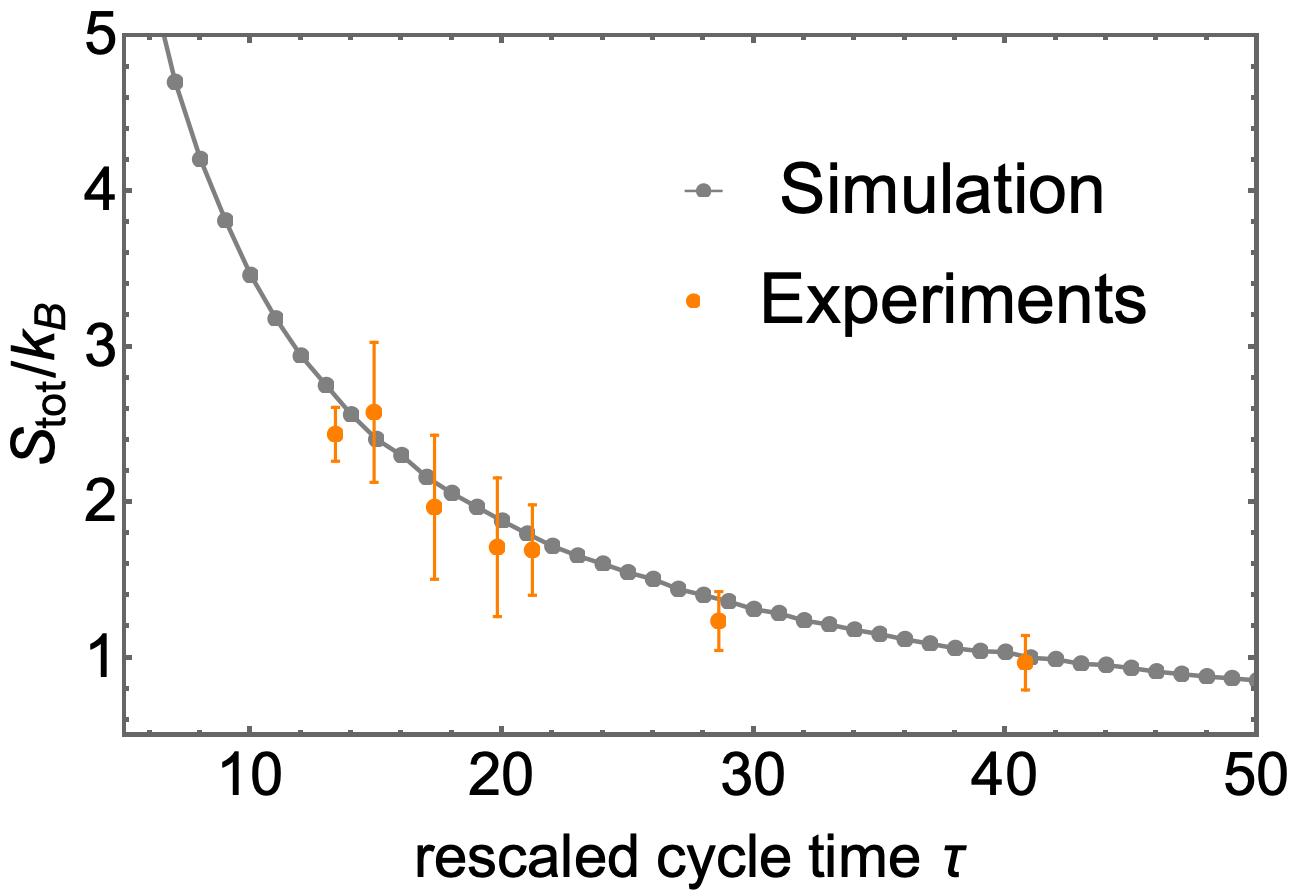}
	\centering
	\caption{\textbf{Inferring entropy production for bit erasure experiment.} Estimated mean total entropy produced $S_{\text{tot}}$ as a function of the rescaled cycle time $t_f/t_m$. We have also performed a comparison with the exact values of $S_{\text{tot}}$ obtained by simulating the Langevin equation with potential $U(x,t)$ given in Eq.~\eqref{BE-eq-2}. The error bars are calculated through least-squares fitting to moments $X_1(t)$ and $X_2(t)$. \bluew{An analysis on calculating the error is given in Eq.~\eqref{error-eqq-BE}.}}
	\label{BE-fig3}
\end{figure}
 
Here, we are interested in estimating the mean entropy produced during the erasure operation by using the bound in Eq.~\eqref{estimate-eq-1}. For this, we take $20-50$ trajectories for different cycle times $t_f$ and measure the first two moments. The precise values of cycle time and number of trajectories considered are given in supplementary table~(II).
We fit the measured moments using the basis of Chebyshev polynomials as
\begin{align}
& X_1(t) = \langle x(t) \rangle = \chi_0 T_0(t) + \chi _1 T_1(t)+....+\chi _s T_s(t), \label{BE-eq-4}\\
& X_2(t) = \langle x^2(t) \rangle = \psi_0 T_0(t) + \psi _1 T_1(t)+....+\psi _s T_s(t), \label{BE-eq-5}
\end{align}
where we typically choose $s=20$. For our analysis, we have rescaled the position by $x \to x / x_m$ and $t \to t/ t _m$ with $t_m = x_m^2 /D$ being the diffusive time scale. We have illustrated the resulting experimental plots in Figure \ref{BE-fig-2}. In the experiment, $E_b /K_B T= 13$, $A =0.2$, $\eta =3$ and $t_m =2.52 ~\text{ms}$. With the form of $X_1(t)$ and $X_2(t)$ in Eqs.~\eqref{BE-eq-4} and \eqref{BE-eq-5}, we use the formula in Eq.~\eqref{estimate-eq-1} to estimate the mean entropy produced as a function of the cycle time.

Figure \ref{BE-fig3} illustrates the primary outcome of this section. The estimated mean entropy production $S_{\text{tot}}$ has been plotted for different dimensionless cycle times $t_f$. The experimental results are essentially in perfect agreement with simulation results despite the very low amount of data and the non-trivial dynamics of the system. This shows that our method is well-suited to study the entropy production of general time-dependent experimental systems.

\bluew{Recall that in estimating the entropy production, we fitted the measured moments using the basis of Chebyshev polynomials in Eqs.~\eqref{BE-eq-4} and \eqref{BE-eq-5}. Due to small number of trajectories, the measured moments are never smooth. Therefore, while carrying out the fitting procedure, we invariably incur some error $\Delta A_2(t) = A_2(t) \Big|_{\text{fit}}-A_2(t) \Big|_{\text{data}}$ and similarly for other quantities. This will, in turn, give rise to error in the inferred value of $S_{\text{tot}}^{\text{12}}(t_f)$. We calculate this error as
%As indicated in Table \ref{traj-table}, for every cycle time, we only have $< 100$ trajectories. Obtaining moments with such few trajectories incur some error, which also gives rise to error in the estimated entropy production in Figure \ref{BE-fig3}. To calculate this error, denoted by $  \Delta S_{\text{tot}}^{\text{12}}(t_f)$, we use the expression in Eq.~\eqref{estimate-eq-1} as follows:  
\begin{equation}
\scalebox{0.91}{$
\begin{split}
\Delta  S_{\text{tot}}^{\text{12}}(t_f) &  = \left| \frac{k_B}{D}\int_0^{t_f} dt ~\left[     \frac{\dot{A}_2(t)}{2 A _2(t) } ~\Delta \dot{A}_2(t)-  \frac{\dot{A}_2(t)^2}{4 A _2(t)^2 } ~\Delta A_2(t) \right.  \right. \\   
& ~~~~~~~\left.  \left.  + 2 \dot{X}_1(t) \Delta \dot{X}_1(t) \right] \right|. 
\end{split}$}
\label{error-eqq-BE}
\end{equation}
Using the values of $\Delta \dot{A}_2(t)$, $\Delta A_2(t)$ and $\Delta \dot{X}_1(t)$, we calculate $\Delta  S_{\text{tot}}^{\text{12}}(t_f)$ via Eq.~\eqref{error-eqq-BE} and take $\pm \Delta  S_{\text{tot}}^{\text{12}}(t_f)  $ as error in the value of the inferred entropy production. This error is then displayed in Figure \ref{BE-fig3}.}

%Our study unravels that the estimated $S_{\text{tot}}$ is well fitted by the function
%\begin{align}
%\frac{S_{\text{tot}}}{k_B} = \frac{\omega}{\tau}
%\end{align}
%where $\omega = 33.83$. In addition to this, 
%We have also compared the estimated $S_{\text{tot}}$ with its exact values obtained by numerically simulating the Langevin equation with parameters corresponding to their experimental values. Quite remarkably, we find that the estimated value is very close to the actual value.

\section*{Conclusion}
\label{con-sec}
In this paper, we have constructed a general framework to derive a lower bound on the mean value of the total entropy production for arbitrary time-dependent systems described by overdamped Langevin equation. The bound, thus obtained, is given in terms of the experimentally accessible quantities, namely the (mixed) moments of the observable. If one only includes the first two moments, one gets a simple analytical expression for the lower bound, whereas including higher moments leads to a numerical scheme. We tested our results both on two analytically solvable toy models and on experimental data for bit erasure, taken from \citep{Bit-2}. The lower bound is close to the real entropy production, even if one only takes into account the first two moments. This makes it a perfect method to infer the entropy production and opens up the possibility to infer entropy production in more complicated processes in future research. One can for example think about biological systems, where upstream processes such as glycolytic oscillations \citep{Glycolytic} or oscillatory dynamics in transcription factors \citep{Transcription}, can effectively lead to time-dependent thermodynamic descriptions. Throughout our paper, we have assumed that the diffusion coefficient is constant, but the results can be extended to spatially-dependent diffusion coefficients as discussed in Supplementary note 7.
%\redw{Furthermore, our method can also be generalised to situations where the diffusion coefficient depends on the position of the particle. This is illustrated in Appendix \ref{appen-spatial-dependence}.}

Compared to some previous works, our work offers several advantages in inferring entropy production. First and foremost, our method manages to infer entropy production from a very limited amount of data. For all the examples discussed in the paper, we obtain a good estimate with of the order of $<1000$ trajectories. For the bit erasure experiment, we actually used $<100$ trajectories [see Supplementary note 6]. On the other hand, for the theoretical example of free diffusion, as demonstrated in Supplementary note 4, the error is approximately $\sim 10\%$ with  $\sim 700$ trajectories. Other proposed methods in the literature typically need at least of the order of $10^4$ trajectories \citep{Otsubo2022} or require data from different measurement set-ups \citep{TUR-6,TUR-7,Koyuk2020}. This has allowed us for the first time, to apply an inference method directly to a time-dependent experimental system and show an excellent agreement between the inferred value and the simulation result. Furthermore, %in  experimental system, we have used a time resolution of $\Delta t = 0.005 \text{s}$, which is not high in comparison to other methods. In general, 
one can easily use smoothing methods to calculate the derivatives of the moments. We therefore anticipate our method to perform well when time-dependent protocols do not change rapidly and the temporal variation in moments is not drastic. %Furthermore, in some cases, we provide exact analytical expressions of the bound which are easy to implement on the available data. This enabled us to avoid any optimisation scheme which is often computationally expensive, making it well-suited for more complex data.

\bluew{Having said this, our method also raises some open questions and a further investigation is needed to answer them. For example, while we have developed our method for continuous overdamped systems, extending these ideas for underdamped systems or discrete-state systems is an interesting future direction. In particular, recently in \citep{Busiello2019}, the connection between the entropy productions in systems governed by the Fokker-Planck equation and those governed by discrete master equations was studied. Exploring the applicability of our method in such a context would be interesting. Moreover, our method gives a tight bound only for the time-dependent systems with conservative forces. For non-conservative forces, although, in principle, our method works, we anticipate it to give a loose bound. In order to achieve a tighter bound, it might be possible to decompose the entropy production into excess and
housekeeping parts arising due to time-dependent driving and non-conservative forcing respectively \citep{GD-1, GD-2, GD-3,GD-4}. Our method might give an estimate of the excess part, while there are proposed methods for the housekeeping part in the steady-state \citep{Estimate-0,Estimate-1, Estimate-2,WTT-1, WTT-2, WTT-3, WTT-4, WTT-5, others-1, others-2, others-3, others-4, others-5, Baiesi-2023}. Combining our method with these methods to infer the entropy production even in presence of non-conservative force fields remains a promising future direction. In some cases, the housekeeping entropy can also be related to an effective system with time-dependent periodic driving \citep{Busiello2018}, a scenario where our method is applicable. Hence, for these cases, it might be possible to use our method even in the non-equilibrium steady-state. Finally, it would be interesting to compare our method with the methodology from \citep{interactions-1, interactions-2}, to infer entropy-production in high-dimensional systems, where it would be impossible to sample the entire phase-space.} We end by stressing that, although we view thermodynamic inference as the main application of our results, the thermodynamic bounds Eqs.~\eqref{estimate-eq-1} and \eqref{gen-eq5} are also interesting in their own rights and might lead to a broad range of applications similar to the thermodynamic speed limit and the thermodynamic uncertainty relations \citep{TUR-1, TSL-4}.

\section*{Data availability}
\noindent
Data available on request from the authors.

\section*{Code availability}
\noindent
Computer codes used in this paper are available on request from the authors.

%\onecolumngrid
\section*{References}
\bibliographystyle{apsrev4-2}
\bibliography{Maindraft.bib}

\section*{Acknowledgement}
We would like to thank Prithviraj Basak and John Bechhoefer for fruitful discussions on the paper and for providing the data used in our study. We also thank Jonas Berx for carefully reading the manuscript. This project has received funding from the European Union’s Horizon 2020 research and innovation program under the Marie Sklodowska-Curie grant agreement No. 847523 ‘INTERACTIONS’ and grant agreement No.~101064626 'TSBC' and from the Novo Nordisk Foundation (grant No. NNF18SA0035142 and NNF21OC0071284).

\section*{Author contributions}
\noindent
K.P. designed research; P.S. and K.P. did the analytical calculations; P.S. implemented numerical codes and simulations. Both authors checked the results and wrote the manuscript.

\section*{Competing interests}
The authors declare no competing interests.

\end{document}